\documentclass[twocolumn,showpacs,preprintnumbers,amsmath,amssymb,prd,letterpaper,floatfix,nofootinbib,superscriptaddress]{revtex4}  
\usepackage{graphicx}
\usepackage{epsfig}
\usepackage{color}
\usepackage{longtable}
\usepackage{hyperref}
\usepackage{latexsym}
\usepackage{amsmath}
\usepackage{amssymb}
\usepackage{dsfont}
\usepackage{verbatim}
\usepackage{bm}

\newcommand{\I}{\mathrm{i}}  
\newcommand{\E}{\mathrm{e}}  
\newcommand{\eq}[1]{(\ref{#1})}
\newcommand{\FC}{\;,}
\newcommand{\FD}{\;.}
\newcommand{\srm}[1]{\mathrm{\scriptstyle #1}}
\newcommand{\qbar}{\overline{q}}
\newcommand{\ubar}{\overline{u}}
\newcommand{\dbar}{\overline{d}}
\newcommand{\ot}{\ensuremath{\frac{1}{2}}}
\newcommand{\be}{\begin{equation}}
\newcommand{\ee}{\end{equation}}
\newcommand{\Npi}{\ensuremath{{N\pi}}}
\newcommand{\unitmatrix}{\mathds{1}}

\begin{document}


\date{\today}
\title{Scattering in the $\pi N$ negative parity channel in
lattice QCD}

\vspace{0.2cm}

\author{C. B. Lang}
\email{christian.lang@uni-graz.at}
\affiliation{Institut f\"ur Physik, FB Theoretische Physik, Universit\"at
Graz, A--8010 Graz, Austria}

\author{V. Verduci}
\email{valentina.verduci@uni-graz.at}
\affiliation{Institut f\"ur Physik, FB Theoretische Physik, Universit\"at
Graz, A--8010 Graz, Austria}

\begin{abstract}
We study the coupled $\pi N$  system (negative parity, isospin $\ot$) based on a
lattice QCD simulation for $n_f$=2 mass degenerate light quarks. Both, standard
3-quarks baryon operators as well as meson-baryon (4+1)-quark operators are
included. This is an exploratory study for just one lattice size and  lattice
spacing and at a pion mass of 266 MeV. Using the distillation method and
variational analysis we determine energy levels of the lowest eigenstates.
Comparison with the results of simple 3-quark correlation studies exhibits
drastic differences and  a  new level appears. A clearer picture of the negative
parity nucleon spectrum emerges. For the parameters of the simulation we may
assume elastic $s$-wave scattering and can derive values of the phase shift.
\end{abstract}


\maketitle

\section{Introduction}

Even if we consider only strong interactions almost all hadrons are unstable.
Calculations in lattice Quantum Chromodynamics (QCD) should therefore take
into account the resonant nature of these states
and the coupled decay channels. The bulk of lattice studies
rely on correlation functions for simple $\qbar q$ or $qqq$-type
operators for mesons or baryons, respectively. Formally one would expect
that in the full quantum field theory with dynamical quarks these simple
meson or baryon operators should (via dynamical vacuum loops) couple to
meson-meson or meson-baryon states. It was found that such
intermediate channels seem to be coupling too weak to be observed (see, e.g.,
\cite{Cohen:2009zk,Mahbub:2009cf,Dudek:2010wm,Bulava:2010yg,Engel:2010my,Engel:2012qp,Mahbub:2012ri}
for  baryon correlation studies). Therefore one needs to include explicitly
hadron-hadron operators  in the set of interpolators, as has been demonstrated
in meson resonance studies
\cite{Lang:2011mn,Aoki:2011yj,Feng:2010es,Lang:2012sv,Mohler:2012na,Pelissier:2012pi,Dudek:2012xn}. 

The interplay between resonance levels and hadron-hadron states has been
discussed in \cite{Luscher:1985dn,Luscher:1986pf,Luscher:1990ux,Luscher:1991cf},
where the resulting energy levels for finite spatial volume were related to the
continuum scattering phase shift in the elastic region. Comparing the energy
levels of a non-interacting hadron-hadron state with those in the case of
interactions one finds a significant level shift (``avoided level crossing'') in
the energy region of the resonance. The effect of such coupled channels depends
on the system parameters. For small volumes and unphysical large quark
masses the two-hadron energy levels may lie high above the observed resonance
levels or - for narrow resonances - outside the influence region of the
resonance.  

Often it is technically not possible (e.g., due to a small volume) 
to determine more than a few lowest energy
levels below the elastic threshold. In the elastic scattering region each energy
level corresponds  to one value of the phase shift and the resonance region 
then cannot be mapped out sufficiently
well. One the other hand, each change of volume or other parameters requires a
completely new simulation  sequence (i.e., generating configurations with
dynamical fermions, quark propagators etc.) Studying interpolators in moving
frames 
\cite{Rummukainen:1995vs,Kim:2005gf,Fu:2011xz,Leskovec:2012gb,Gockeler:2012yj}
allows to obtain further values on the same configurations. Unfortunately, for
coupled channels with two hadrons of different mass, there can be mixing between
different partial waves, complicating the situation. Another complication is the
opening of inelastic channels.

Starting from continuum models for a scattering process, based
on phenomenologically determined parameters,
one can also derive the energy levels on finite volume lattices
\cite{Bernard:2008ax,Doring:2011ip,Roca:2012rx,Hall:2012wz}. Coupled channel
potential models or Unitarized Chiral Perturbation Theory motivated models in
that way allow to compare with lattice results. Alternative methods to identify
resonance parameters have been discussed in that context
\cite{Bernard:2008ax,Meissner:2010rq,Giudice:2012tg}. 

A particularly interesting case is the negative parity nucleon channel. There we
have two low lying resonances $N^*(1535)$ and $N^*(1650)$ which couple to $\Npi$
in $s$-wave.  Above the 10\% level there are also further inelastic decays 
$N^*(1535)\to N\eta$ and $N^*(1650)\to N\eta, \Lambda K$. So far lattice
simulations of this channel, that have determined ground state energy levels and
further excitations, included only 3-quark interpolators
\cite{Bulava:2010yg,Engel:2010my,Engel:2012qp,Mahbub:2012ri}. 
In these studies two low lying
energy levels have been identified and assigned to the two negative parity
resonances. However, the lower of the two levels showed a tendency to lie below
the $N^*(1535)$.

In order to clarify the situation we study here for the first time the coupled
system of 3-quark nucleon interpolators and pion-nucleon interpolators in the
negative parity channel. The calculation requires the computation of many more
correlation graphs than before, including the notoriously demanding backtracking
quark line contributions. We therefore use the distillation method
\cite{Peardon:2009gh} for determining the cross correlation matrix for up to 9
interpolators. We use gauge configurations with $n_f=2$ mass degenerate
dynamical quarks (of improved Wilson type) with a pion mass of 266 MeV. The
$16^3\times 32$ lattices have spatial extent of 1.98 fm with
$L m_\pi\approx 2.68$ (for details see
Table \ref{tab:gauge_configs}). The energy levels are
obtained with the variational method
\cite{Michael:1985ne,Luscher:1985dn,Luscher:1990ck,Blossier:2009kd}.

The energy levels of the eigenstates in case of a finite spatial extent $L$ are discrete. They are
determined by diagonalizing the correlation matrix of interpolating operators.
The set of these interpolators has to be large enough to allow the
representation of the eigenstates. For total momentum zero the pion-nucleon
operators will have the form $N(n)\pi(-n)$ where $n$ abbreviates the possible
quantized momentum  values $2n\pi/L$. For the non-interacting situation the
corresponding energies are straightforward to compute, for the interacting case
they are  shifted and have to be determined numerically from the correlation
matrix. We need to consider all interpolators that may couple to the system in
the energy region where one expects eigenstates. Obviously the 3-quark
interpolators and the interpolator $N(0)\pi(0)$ have to be included. In our
setting already the $s$-wave operator $N(1)\pi(-1)$ lies high above the lowest
energy level. The same holds for a possible $N\eta_2$ channel (note, that for
only two dynamical quarks there is just one $\eta$ meson called $\eta_2$).  We
find that the spectrum shows a clear difference whether the pion-nucleon operator
is included or not. If the pion-nucleon interpolator is included 
we observe one more level below threshold, typical for
attractive channels, and the next two levels are shifted closer to the expected
resonances. 

Following Sect. \ref{Sec_Methods} where we present the parameters and methods
used, we discuss the results in Sect. \ref{sec:results}. The appendix lists the
necessary Wick contractions for the meson-baryon correlators.
\section{Methods}
\label{Sec_Methods}

\subsection{Lattice action and configurations}

We use configurations from the study of re-weighting techniques 
\cite{Hasenfratz:2008ce,Hasenfratz:2008fg} generously provided by the authors.
The gauge configurations were generated for $n_f= 2$ flavors of mass-degenerate
light quarks and a tree level improved Wilson-Clover action  with gauge links
smeared using one level of normalized hypercubic smearing (nHYP smearing). The
valence u/d quarks have the same mass as the sea u/d quarks. Table
\ref{tab:gauge_configs}  lists the parameters used for the simulation along with
the number of (approximately independent) gauge configurations used, the lattice
spacing, volume and the pion mass (for details see
\cite{Lang:2011mn,Lang:2012sv}). We note that the small value 
$L m_\pi\approx 2.68$
may lead to finite size effects which we cannot identify in this study, since we
have just one lattice size available.

\begin{table}[t]
\begin{ruledtabular}
\begin{tabular}{ccccccc}
$N_S^3\times N_T$ & $\beta$ & $a$[fm] & $L$[fm] & $L\,m_\pi$ &\#configs & $m_\pi$[MeV]\\ 
\hline
$16^3\times32$ & 7.1 & 0.1239(13) & 1.98 & 2.68 & 280 & 266(3)(3) \\
\end{tabular}
\end{ruledtabular}
\caption{\label{tab:gauge_configs} Configurations used for the present study.
$N_S$ and $N_T$ denote the number of lattice points in spatial and time
directions, and $L=N_S a$ is the size in physical units.}
\end{table}

\subsection{Determination of energy levels}

Due to the finiteness of the spatial volume, the energy spectrum of the
correlation functions is discrete.  We determine the energy levels of the $N$
and the $N\,\pi$ system with the variational method
\cite{Michael:1985ne,Luscher:1985dn,Luscher:1990ck,Blossier:2009kd}. For a given
quantum channel one measures the Euclidean cross-correlation matrix $C(t)$
between several interpolators,
\be
C_{ij}(t)=\langle O_i(t) \,\overline O_j(0)\rangle
=\sum_n\langle O_i(t)|n\rangle\E^{-E_n t}\langle n|\overline O_j(0)\rangle\FC
\ee
where the operators are located on the corresponding Euclidean time slices. The
generalized eigenvalue problem 
$C(t)\vec u_n(t)=\lambda_n(t)C(t_0)\vec u_n(t)$ disentangles the
eigenstates $|n\rangle$ with the 
eigenvalues
\begin{equation}
\lambda_n(t,t_0)= \E^{-E_n (t-t_0)} 
\left(1+ \mathcal{O}\left(\E^{-\Delta E_n (t-t_0)} \right) \right)\FC
\end{equation}
where $\Delta E$ may be as small as the distance to the next nearby energy
level. From the exponential decay one determines the energy values of the
eigenstates by exponential fits. The stability of the eigenvectors with regard to $t$ and
the so-called effective energies 
\begin{equation}\label{eq:effenergy}
E_n(t)=\log \frac{\lambda_n(t)}{\lambda_n(t+1)}
\end{equation}
indicate the suitable fit range by exhibiting plateau-like behavior. 
The set of interpolators should be large enough to allow the system to reproduce
the physical eigenstates. Neglecting important interpolators may  obscure the
result. On the other hand, in the calculations it is not  possible to have a
complete set of interpolators  and one is limited to a reasonable subset. Also,
the statistical quality of $C(t)$ is an issue. The reliability of the obtained 
energy levels decreases  for higher $|n\rangle$, with the ground state being the
most  reliable one. 

The energy
values are extracted using correlated fits of $\lambda_n(t)$ to one and two
exponentials.  A possible source of systematic error is the choice of the fit range in $t$. From the
effective energy plots  (cf., Figs.~\ref{fig:posparnucleon} and
\ref{fig:doubleplot}), the range  of stability of the eigenvectors and the
$\chi^2$ dependence of the fits we estimate suitable fit ranges. 
The two exponential fits start at smaller $t$ and we verify that
the extracted levels agree with  results obtained from one-exponential fits
starting at larger $t$.

\subsection{Interpolators}\label{subsec:interpolators}

The $\Npi$ system can be projected to isospin $\ot$ and $\frac{3}{2}$,
experimentally accessible by $\pi^\pm p$ scattering. Here we study only the
isospin $\ot$ sector.

For the charged nucleon interpolator we use the operator (on a given time slice)
\begin{align}\label{eq:defN}
&(N_\pm^{(i)})_\mu(\vec p=0)=\nonumber\\
&\qquad\sum_{\vec x}\epsilon_{abc}\, \left(P_\pm\,\Gamma_1^{(i)}\, u_a(\vec x)\right)_\mu\, 
\left( u_b^T(\vec x)\, \Gamma_2^{(i)}\, d_c(\vec x)\right)\FC
\end{align} 
and for the neutral one with the quarks $d,u,d$. $(\Gamma_1, \Gamma_2)$ can
assume the three values $(\unitmatrix,C\gamma_5)$, $(\gamma_5,C)$ and
$(\I\unitmatrix,C\gamma_t\gamma_5)$ for $i=1,2,3$. $C$ denotes the charge
conjugation matrix, $\gamma_t$ the Dirac matrix in time direction, and
$P_\pm=\ot(1\pm\gamma_t)$ the parity projector. We sum over all points of the
time slice in order to project to zero momentum. Summation over the color
indices $a, b, c$ and the (not shown) Dirac indices is implied. 

In the
distillation approach  (see Subsect. \ref{subsect:distillation} below) the
sources are smeared  combining  $N_v$ eigenvectors.  For the nucleon 3-quark
interpolators we choose $N_v=32$ and $N_v=64$ and thus with the three different
Dirac structures have six operators.

The pion interpolators read
\begin{align}\label{eq:defpi}
\pi^+(\vec p=0)&=\sum_{\vec x} \dbar_a(\vec x) \gamma_5 u_a(\vec x)\FC\nonumber\\
\pi^0(\vec p=0)&=\sum_{\vec x} \frac{1}{\sqrt{2}} 
\left( \ubar_a(\vec x) \gamma_5 u_a(\vec x)
-\dbar_a(\vec x) \gamma_5 d_a(\vec x) \right)  \FC
\end{align}
where summation over the color index $a$ is implied.

We consider the $\Npi$ system in the rest frame. The leading $s$-wave
contribution then comes from the interpolator with both particles at rest,
\be\label{eq:defNpi}
\Npi(\vec p=0)=\gamma_5 N_+(\vec p=0)\pi(\vec p=0)\FC
\ee
where $N_+$ denotes the positive parity nucleon and the factor $\gamma_5$ 
ensures negative parity for the interpolator. In the distillation approach  we
choose for the $\Npi$ channel  $N_v=32$ and thus with the three different
nucleon interpolators have three operators.

We project to isospin $\ot$ by choosing the combination
\be
O_\Npi(I=\ot,I_3=\ot) = p \pi^0 + \sqrt{2}\, n \pi^+\FC
\ee
with $p$ and $n$ denoting the charged and the neutral nucleon  according to
\eq{eq:defN}.

The negative parity $N^*$ channel becomes quickly inelastic (see, e.g.,
\cite{Arndt:2006bf,Manley:1992yb,Koch:1985bn,Cutkosky:1979fy}). According to the
Particle Data Group \cite{Beringer:1900zz} the main decay channel is $\Npi$
(35-55\% for $N^*(1535)$, 50-90 \% for $N^*(1650)$). The second largest decay
rate is to $N\eta$ ( 42$\pm$ 10\% for $N^*(1535)$, 5-15 \% for $N^*(1650)$).
Most of the rest of 10-20 \% is $\Npi\pi$ and, for $N^*(1650)$,  also $\Lambda
K$.  For a lattice calculation at physical quark masses one would need to
include the inelastic channels. This is beyond present days capacities.

In our case ($n_f=2$) there is just one pseudoscalar meson $\eta$ called
$\eta_2$, with a mass larger than 800 MeV \cite{McNeile:2007fu,Jansen:2008wv}.
With our parameters (see Sect. \ref{sec:results}) these inelastic
channels would thus have thresholds above 1900 MeV. The lowest state with total
momentum zero but non-zero relative momentum $N(1)\pi(-1)$ (momentum units
$2\pi/(16 a)$) has a (non-interacting) energy of 1920 MeV, as well. These energy
values are above the observed lowest three levels. We cannot exclude that in
particular the highest of these may be influenced by the $N(1)\pi(-1)$ state.

\subsection{Distillation method and correlation function}\label{subsect:distillation}

We compute the correlation matrix entries with help of the distillation
method  \cite{Peardon:2009gh}. This method has been successfully applied in
several studies, including baryon correlation functions 
\cite{Dudek:2010wm,Bulava:2010yg,Dudek:2011tt,Lang:2011mn,Liu:2012ze,Lang:2012sv,Mohler:2012na}.
It also allows for a reliable evaluation of the partially disconnected
diagrams. On a given time slice one introduces separable (i.e., expressed by
a sum of products separating the dependence on $\vec x$ and   $\vec x'$)
quark smearing sources in the form
\begin{align}
q_{c,\alpha}(\vec x)& \to \sum_{\vec x'}S_{cd}(\vec x,\vec x') q_{d,\alpha}(\vec x') \nonumber\\
&\equiv \sum_{\vec x'}\sum_i^{N_v} v^{i}_c(\vec x) v^{i*}_d(\vec x') q_{d,\alpha}(\vec x')\FC
\end{align}
where  $c,d$ and $\alpha$ denote color and Dirac indices and summation over the
color indices is implied. A suitable choice for the $v^{i}$ is the
eigenvectors of the spatial lattice Laplacian \cite{Peardon:2009gh}. Summing
over all eigenvectors reproduces the delta function, the spectral representation
of unity.  In actual calculation one truncates the sum and uses the lowest 
eigenmodes or subsets. The value of $N_v$ depends on the lattice size and $N_v$
between 32 and 96 was found suitable for our situation \cite{Lang:2011mn}.

The advantage of the distillation approach lies in its versatility.  Instead of
quark propagators  $G_{c\mu;d\nu}(x,x_0)$ from one source located in $x_0$ to
other points on the lattice one now computes propagators between eigenmode
sources, so-called perambulators
\begin{align}\label{eq:deftau}
&\tau_{\mu\nu}(j,t_\srm{snk};i,t_\srm{src})
\equiv \nonumber\\
&\qquad\qquad
\sum_{\vec x,\vec y,c,d}v^{j*}_d(\vec x,t_\srm{snk}) G_{d\mu;c\nu}(\vec x,\vec y) v^{i}_c(\vec y,t_\srm{src})
\FD
\end{align}
The interpolator structure decouples from the calculation of the perambulators
completely. E.g., meson correlators assume the form
\begin{align}
C(t_\srm{snk},t_\srm{src})=\,&\langle M(t_\srm{snk}) M^\dagger(t_\srm{src})\rangle
\nonumber\\
=\,&\phi_{\mu\nu}(n,k;t_\srm{src})  \tau_{\nu\alpha}(k,t_\srm{src};i,t_\srm{snk})\nonumber\\
&\phi_{\alpha\beta}(i,j;t_\srm{snk})\tau_{\beta\mu}(j,t_\srm{snk};n,t_\srm{src})\FC
\end{align}
where $M$ denotes a meson interpolator like, e.g., 
the pion of Eq. (\ref{eq:defpi}) and summation over the source index ($i,j,k,n$) pairs and the Dirac index
($\alpha,\beta,\mu,\nu$) pairs is implied. Due to $\gamma_5$-hermiticity
of the Dirac operator the perambulator for sink to source
can be expressed by that from source to sink,
\be
\tau_{\nu\alpha}(k,t_\srm{src};i,t_\srm{snk})
=\gamma_{5,\alpha\alpha'} 
\tau^\dagger_{\alpha'\nu'}(i,t_\srm{snk};k,t_\srm{src})\gamma_{5,\nu'\nu}\FD
\ee 
The meson interpolator type
is specified in
\begin{align}
\phi_{\alpha\beta}(i,j;t)&=D_{\alpha\beta} \;
\sum_{\vec x,\vec y} v^{i*}_d(\vec y) F_{dc}(\vec y,\vec x)  v^{j}_c(\vec x)\nonumber\\
&\equiv D_{\alpha\beta} \;\widehat\phi(i,j;t)\FD
\end{align}
The factors $D$ and $F$ represent the Dirac structure and momentum projection or
derivative terms related to the quantum numbers of the meson. Only $\phi$ has to
be recomputed  for each meson interpolator whereas the perambulator remains the
same. 

For 3-quark interpolators like the baryons one obtains contributions to the
correlation function of the form
\begin{align}
C_{\mu\nu}(t_\srm{snk},t_\srm{src})&=\langle N_\mu(t_\srm{snk}) \overline N_\nu(t_\srm{src})\rangle\nonumber\\
&=\phi_{\mu\alpha\beta\gamma}(i,j,k;t_\srm{snk})\nonumber\\
&\qquad \tau_{\alpha\alpha'}(i,t_\srm{snk};i',t_\srm{src})\nonumber\\
&\qquad \tau_{\beta\beta'}(j,t_\srm{snk};j',t_\srm{src})\nonumber\\
&\qquad \tau_{\gamma\gamma'}(k,t_\srm{snk};k',t_\srm{src})\nonumber\\
&\qquad\phi_{\nu\alpha'\beta'\gamma'}^{\dagger}(i',j',k';t_\srm{src})\FD
\end{align}
For an interpolator $N$ (without derivatives) $\phi$ assumes the form
\begin{align}
\phi_{\nu\alpha\beta\gamma}(i,j,k;t)&=
D_{\nu\alpha\beta\gamma}\;\sum_{\vec x}\epsilon_{abc} 
v^{i}_a(\vec x)
v^{j}_b(\vec x)
v^{k}_c(\vec x)
F(\vec x)
\nonumber\\
&\equiv
D_{\nu\alpha\beta\gamma}\widehat \phi(i,j,k;t)
\FD
\end{align}
Again, $D$ carries the Dirac structure and $F$ the possible total momentum projection
factors.

We also project the correlation functions to definite parity with the
projection  operators $P_\pm=\ot(1\pm \gamma_t)$. In App. \ref{app:contractions}
we list the necessary contraction  terms expressed in terms of the
perambulators.

\subsection{Energy levels: interpretation}\label{subject:levelsinterpretation}

We study the $\Npi$ system in the $\ot^-$ channel in $s$-wave in the rest
frame.  From the energy value
\be\label{eq:energy}
E=\sqrt{s}=\sqrt{(p_N+p_\pi)^2}=\sqrt{p^{*2}+m_\pi^2}+\sqrt{p^{*2}+m_N^2}
\ee
we extract the momentum $p^{*}=|\mathbf{p}^*|$ with
\begin{equation}\label{eq:pstar_q2}
p^{*2}=\frac{[s-(m_N+m_\pi)^2][s-(m_N-m_\pi)^2]}{4s}\FC
\end{equation}
and the dimensionless product of the momentum  and the spatial lattice size
\be\label{eq:def_of_q}
q=p^*\frac{L}{2\pi}\FD
\ee

\begin{figure}[t]
\begin{center}
\includegraphics[width=0.4\textwidth,clip]{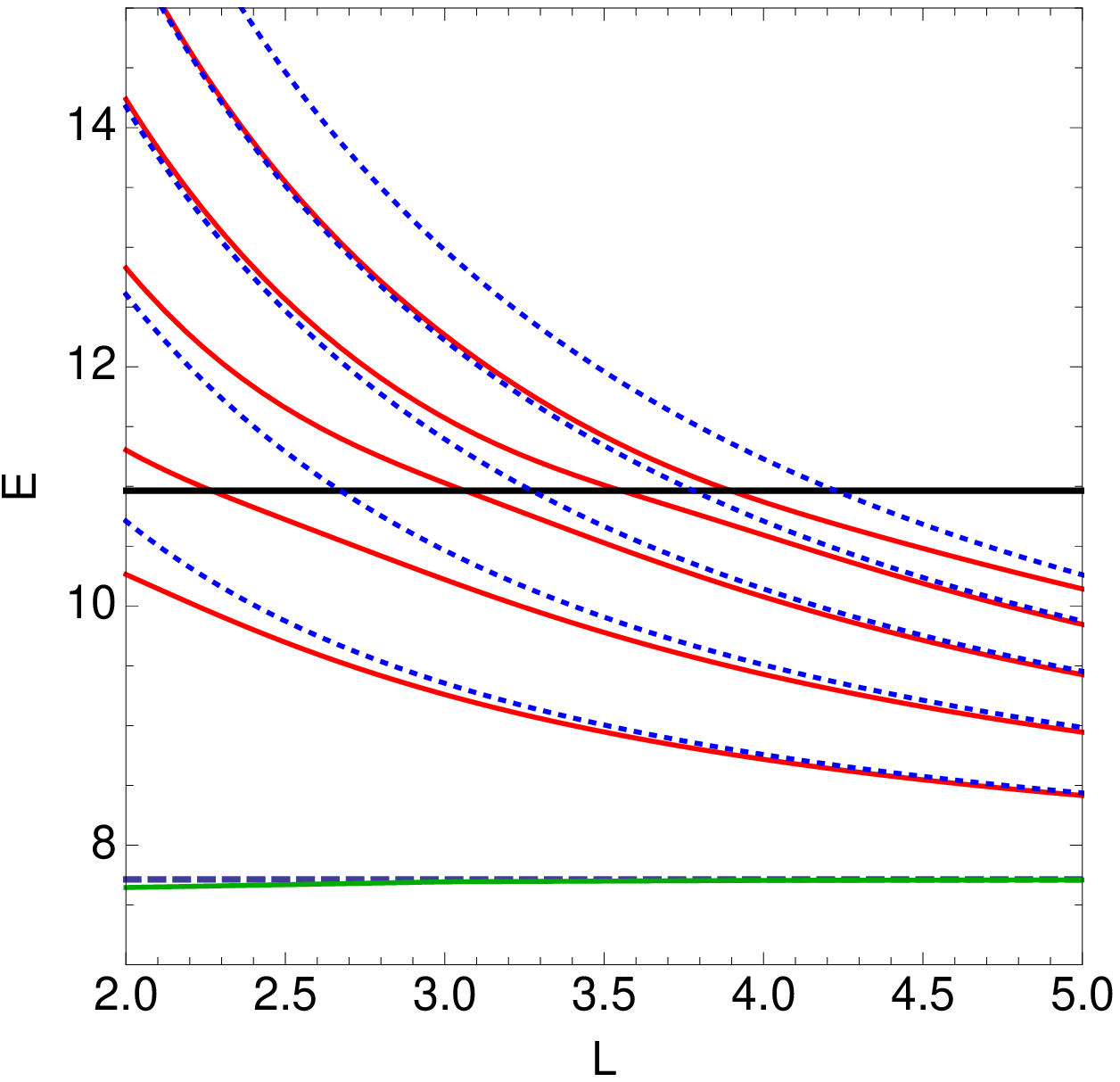}\\
\includegraphics[width=0.4\textwidth,clip]{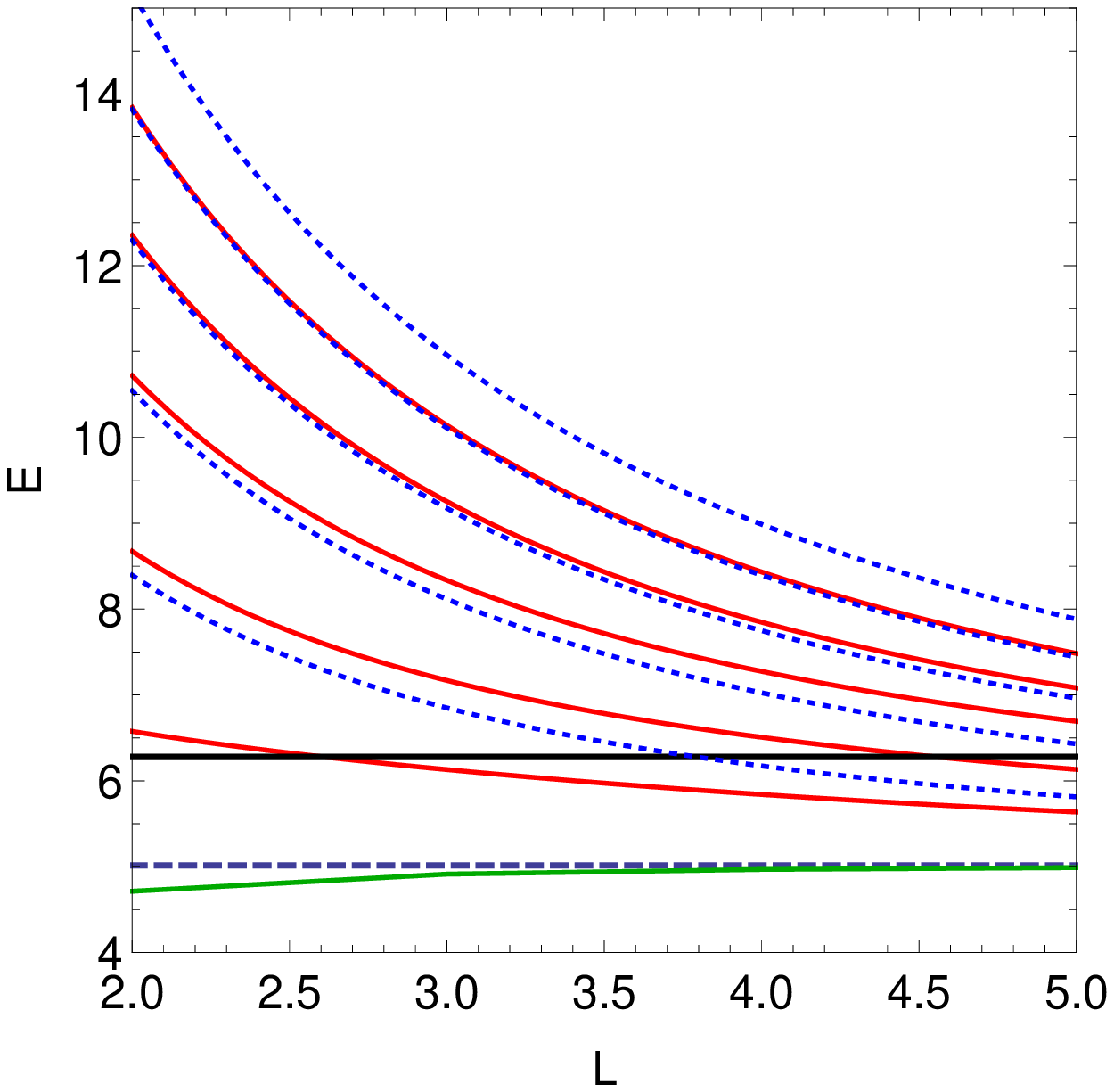}
\end{center}
\caption{Lowest energy levels vs. the spatial lattice size (both in units of the
pion mass). Upper plot:  physical pion, nucleon and $N^*(1535)$ masses,
comparing the non-interacting levels (dotted) with the levels distorted due to
interaction (full lines); the broken horizontal line indicates the threshold,
the horizontal thick line the $N^*(1535)$-mass; the $N^*$ is parametrized as an
elastic resonance with a decay width of 150 MeV. Lower plot: Unphysical values
for $m_\pi=266$~MeV, $m_N =1068$~MeV; for $N^*$ the mass is chosen as 1670 MeV
without changing the coupling strength. In both cases the lowest possible state
is $N(0)\pi(0)$,  which coincides with the threshold in the non-interacting
case. For attractive interaction the level moves slightly below  the threshold
to negative $q^2$.}\label{fig:energylevels}
\end{figure}

For a system of non-interacting pions and nucleons the energy levels for given
lattice size can be straightforwardly computed (dotted lines in 
Fig.~\ref{fig:energylevels}). For the interacting case, with localized interaction
region and  in the elastic domain, L\"uscher 
\cite{Luscher:1985dn,Luscher:1986pf,Luscher:1990ux,Luscher:1991cf} has given a
relation between energy levels and phase shift,
\begin{equation}\label{zeta}
\tan \delta(q)=\frac{\pi^{3/2} q}{\mathcal{Z}_{00}(1;q^2)}\FC
\end{equation}
where $q$ is given in \eq{eq:def_of_q}, and the generalized zeta function
$Z_{lm}$ is given in \cite{Luscher:1990ux}.

Assuming a phase shift parameterization, one can numerically invert that
relation and obtain the modified energy levels, which exhibit the phenomenon of
avoided level crossing by level ``transmutation''. 

The $s$-wave amplitude may be written
\be
T=\E^{\I\delta}\sin \delta=\frac{1}{\cot \delta-\I}\FD
\ee
We also define for convenience
\be\label{eq:rho0}
\rho_0(s)=\frac{p^*}{\sqrt{s}} \cot \delta = 
\frac{2 Z_{00}(1;q^2)}{L \sqrt{s\pi}}\FC
\ee
with the effective range parameterization near threshold
\be\label{eq:scattlength}
\sqrt{s}\,\rho_0(s)=\frac{1}{a_0}+\mathcal{O}(p^{*2})\FC
\ee
and scattering length $a_0$.
If the first resonance is of Breit-Wigner shape, then $\rho_0$ 
can be approximated linearly,
\be\label{eq:BW}
\rho_0(s)=\frac{1}{\gamma}(s_R-s)\FD
\ee
Here $s_R$ denotes the resonance position and $\gamma$ is related to the width 
\be\label{eq:width}
\Gamma=\frac{p^*(s_R)}{s_R} \gamma
\ee
or the coupling constant $\gamma=g^2/{6\pi}$.

The $N(\ot^-)$ ($s$-wave) scattering amplitude is shown in the data analysis of 
\cite{Arndt:2006bf} and has an intricate behavior, becoming quickly inelastic.
In a simplification of that case let us study the situation with just one
elastic resonance. In that case the phase shift and elastic amplitude can be
modeled where we have used a resonance mass of 1535 MeV and a width of 150 MeV.
The resulting energy levels demonstrating the expected avoided level crossing
are also shown in  Fig.~\ref{fig:energylevels}.

In the lower part of  Fig. \ref{fig:energylevels} we show the situation where the
pion mass has the larger value 266 MeV. (Note that it is also used as unit mass
in that plot.) The values of the stable nucleon has been set to 1068 MeV and the
resonance position to 1670 MeV, all values close to the results of our
calculation  to be discussed in Sect. \ref{sec:results}. The coupling strength
$\gamma$ at the resonance position is unchanged.

In this setting the picture changes drastically. For our lattice size we have
$L\approx 2.68$ (in units of $m_\pi$) and the  energy level $N(1)\pi(-1)$ lies
clearly above the resonance. The lowest possible state is $N(0)\pi(0)$, which
coincides with the threshold in the non-interacting case. For attractive
interaction the level moves slightly below the threshold to negative $q^2$,
which is a finite volume artifact.

Choosing interpolators with non-zero total momentum (``moving frame'') allows in
principle to obtain  further energy levels and thus additional values of the
phase shift. For the case of two particle of equal mass this was discussed in
\cite{Rummukainen:1995vs,Kim:2005gf} and has been used in various studies of the
$\pi\pi$-system. The situation for pairs of  hadrons with different masses is
more complicated \cite{Fu:2011xz,Leskovec:2012gb,Gockeler:2012yj} since there
even and odd partial waves may mix. In this study we rely on the case of zero
momentum. Smaller quark masses will require the consideration of further $\Npi$
operators and other interpolators.

L\"uscher's relation holds in the elastic region. Most often inelasticity sets
in early due to coupled channels. An alternative approach is the
inverse procedure, starting with a (unitarized) coupled channel parameterization
of the scattering matrix in continuum and then determining the expected discrete
energy levels on finite volumes, see, e.g.,
\cite{Bernard:2008ax,Doring:2011ip,Roca:2012rx,Hall:2012wz}. The lattice results
for the energy levels can then be interpreted along these lines.

\section{Results}
\label{sec:results}
\subsection{Pion and nucleon, non-interacting}

The masses of the free pion and the ground state nucleon $N(\ot^+)$ have to be
estimated with the highest possible precision in order to perform the subsequent
analysis. For the gauge  configurations used here the pion was studied carefully
in  \cite{Lang:2011mn,Lang:2012sv} where the value $a\,m_\pi=0.1673(2)$ was
obtained and we use this value here as well.

\begin{figure}[t]
\begin{center}
\includegraphics[width=0.48\textwidth,clip]{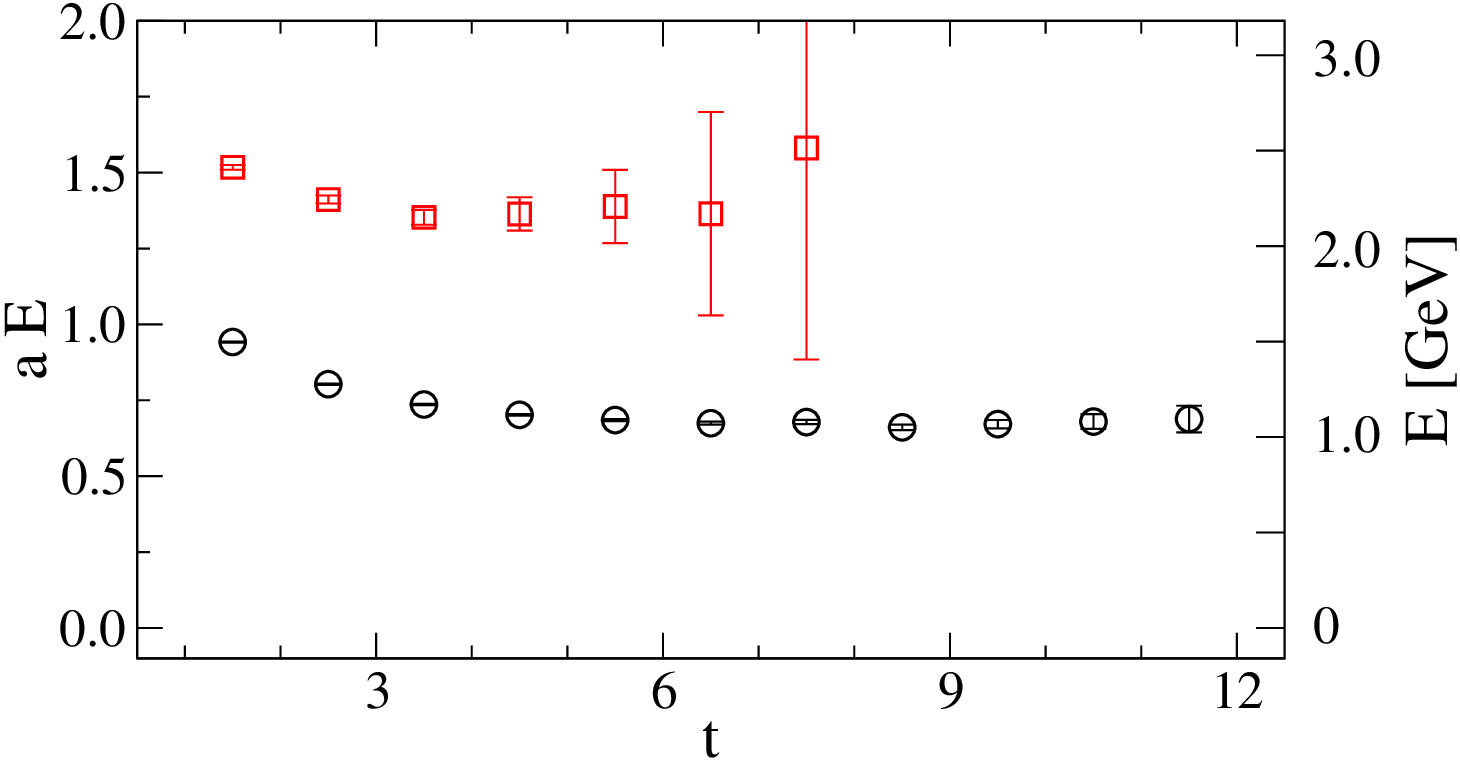}\\
\end{center}
\caption{The effective energy values for the $N(\ot^+)$ channel (with
3-quark interpolators).}\label{fig:posparnucleon}
\end{figure}

For the positive parity nucleon $N(\ot^+)$ we study the correlation matrix for
the six operators 
\begin{align}
\mathcal{O}^+_1,\,\mathcal{O}^+_2,\,\mathcal{O}^+_3 
&= N_+^{(1)},\,N_+^{(2)},\,N_+^{(3)} &\textrm{~with~} &N_v=32\FC\nonumber\\
\mathcal{O}^+_4,\,\mathcal{O}^+_5,\,\mathcal{O}^+_6 
&= N_+^{(1)},\,N_+^{(2)},\,N_+^{(3)} &\textrm{~with~} &N_v=64\FD
\end{align}
The correlation matrix is analyzed with the variational method as discussed
above. The ground state shows a stable plateau behavior in the effective energy
plot  Fig.~\ref{fig:posparnucleon}. The first excitation is considerably higher
than the expected Roper resonance. This observation is shared by other recent
studies (see, e.g., \cite{Cohen:2009zk,Bulava:2010yg}) but 
disputed \cite{Mahbub:2009cf,Mahbub:2010rm}. The reason for the high value may
lie in the incompleteness of the interpolator basis, i.e., possibly missing
important 5-quark interpolators. To solve this puzzle is not the object of  our
study. Our value of the ground state nucleon (fit range 6-12) is
$a\,m_N=0.672(4)$ (corresponding to $m_N=1068(6)$ MeV).

\subsection{Interacting $\Npi$ system}

We compute the full correlation matrix for the following operators:
\begin{align}\label{eq:operators}
\mathcal{O}^-_1,\,\mathcal{O}^-_2,\,\mathcal{O}^-_3 
&= N_-^{(1)},\,N_-^{(2)},\,N_-^{(3)} &\textrm{~with~}&N_v=32\FC\nonumber\\
\mathcal{O}^-_4,\,\mathcal{O}^-_5,\,\mathcal{O}^-_6 
&= N_-^{(1)},\,N_-^{(2)},\,N_-^{(3)}&\textrm{~with~}&N_v=64\FC\nonumber\\
\mathcal{O}^-_7,\,\mathcal{O}^-_8,\,\mathcal{O}^-_9 
&= O_\Npi^{(1)},\,O_\Npi^{(2)},\,O_\Npi^{(3)}&\textrm{~with~} &N_v=32\FC
\end{align}
with the definition from \eq{eq:defN} and \eq{eq:defNpi}.

\begin{figure}[t]
\begin{center}
\includegraphics[width=0.48\textwidth,clip]{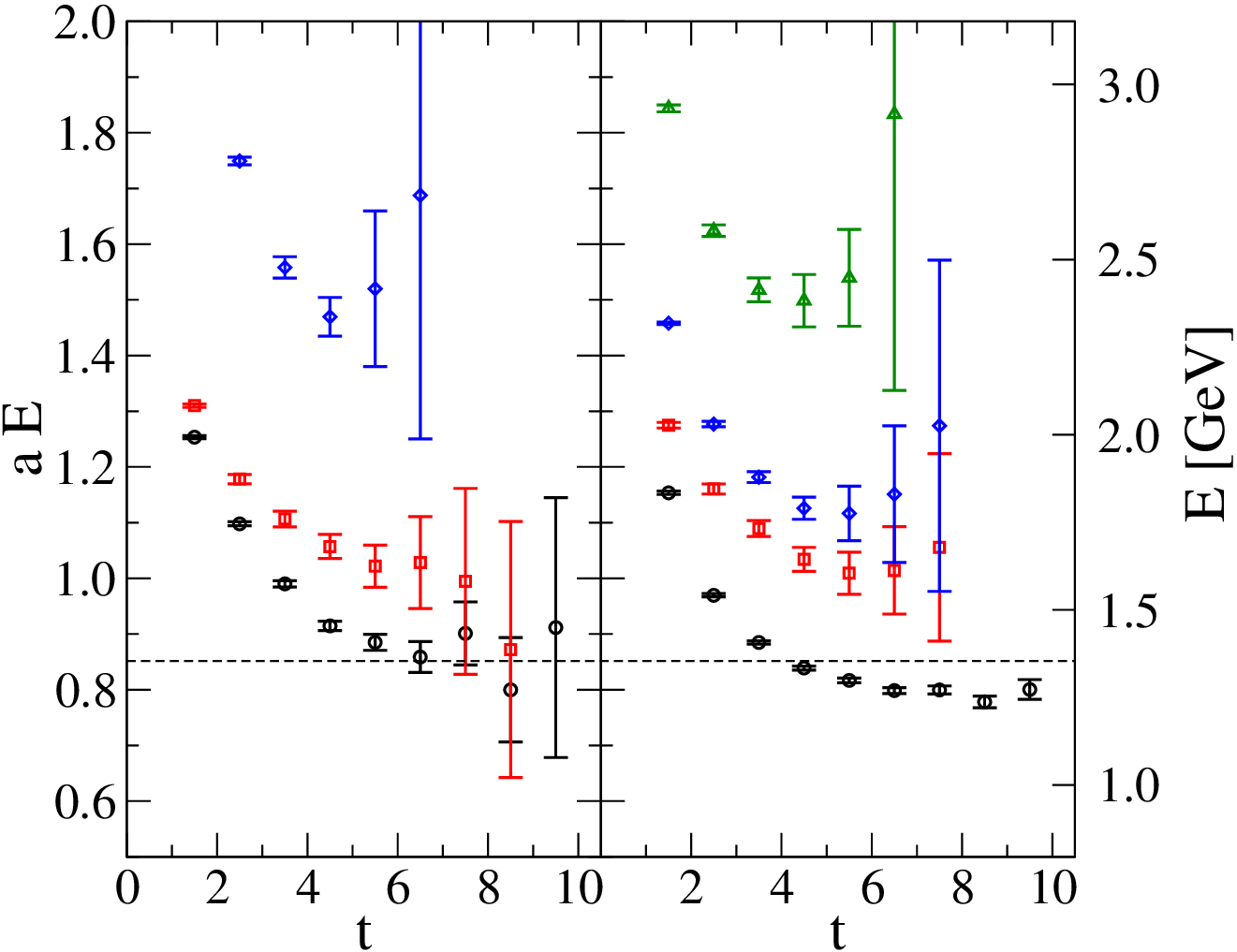}
\end{center}
\caption{Left: effective energy values for the case without $\Npi$ contribution,
right: including $\Npi$ interpolators. The horizontal broken line indicates the
threshold value $m_N+m_\pi$.}\label{fig:doubleplot}
\end{figure}

\begin{figure}[t]
\begin{center}
\includegraphics[width=0.4\textwidth,clip]{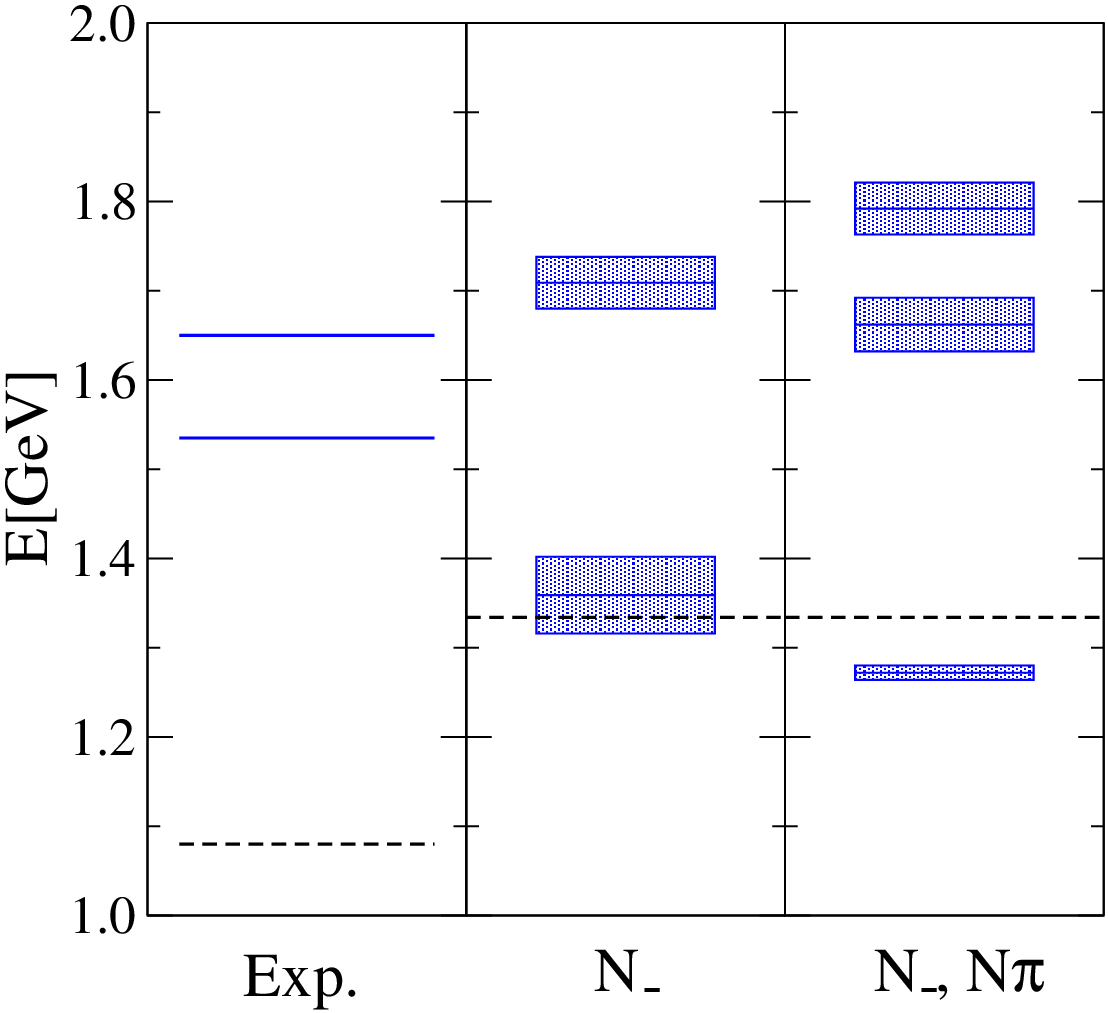}
\end{center}
\caption{Comparison of the energy levels. Left: physical mass values
(experiment). Middle: result when using only 3-quark interpolators. Right:
result when pion-nucleon interpolators are included. The dashed lines indicate
the scattering thresholds.}\label{fig:comparison}
\end{figure}

Let us first consider results for the subset of 3-quark interpolators
$\mathcal{O}^-_1$-$\mathcal{O}^-_6$. It turns out that inclusion of the type
$N_-^{(3)}$ does not improve the quality of the diagonalization results. We
therefore use only the subset
$(\mathcal{O}^-_1,\mathcal{O}^-_2,\mathcal{O}^-_4,\mathcal{O}^-_5)$. We reproduce the
usual (see, e.g. \cite{Engel:2010my,Bulava:2010yg})  pattern of energy levels
(see left hand plot of Fig.~\ref{fig:doubleplot}), which have been assigned to the 
two $N^*$ resonances. However, as has been observed in \cite{Engel:2013ig},
towards smaller pion masses the lower level moves 
close to the expected threshold and thus lies unexpectedly low if compared to
the $N^*(1535)$. 
The situation is shown in Fig.~\ref{fig:comparison} (middle). The energy levels have the values 1.359(43) GeV
(exponential fit, fit range 6-10) and 1.709(29) GeV (fit range 4-9).

This picture changes significantly, when one includes the $\Npi$-interpolators
in the correlation matrix. The right hand plot of Fig.~\ref{fig:doubleplot}
shows the effective energy levels when using operators
$\mathcal{O}^-_1,\mathcal{O}^-_2,\mathcal{O}^-_4,\mathcal{O}^-_5,\mathcal{O}^-_7,\mathcal{O}^-_8,\mathcal{O}^-_9$
in the analysis. The exponential fits to the corresponding eigenvalues and the
resulting energy levels are listed in Table \ref{tab:results}.

\begin{table*}[t]
\begin{ruledtabular}
\begin{tabular}{cccllllll}
level  &$t_0$      & fit      & $a\,E_n=a\,\sqrt{s}$ &$E=\sqrt{s}$ & $\tfrac{\chi^2}{d.o.f.}$  & $a\,p^*$  & $\rho_0$&$\delta$  \vspace{-3pt} \\    
   $n$ &              & range&                                 & [GeV]      &                           &                          &                & [degrees] \\
   \hline
1        &           1  & 6-12  & 0.800(5)                   &1.272(8)     &   6.12/5          & 0.0985(57) $\I$ &  0.149(48) & 68(59)$\I$\\
2        &           1  & 4-8   & 1.045(19)                 & 1.662(30)    &   2.46/3          & 0.2726(155)     &  0.007(42) & 89(9)       \\
3        &           1  & 4-8   & 1.127(18)                & 1.792(29)     &   0.67/3          & 0.3362(134)     &  0.279(108)& 47(10)     \\
\hline
\end{tabular}
\end{ruledtabular}
\caption{ Final results for the lowest three energy levels of the coupled $\Npi$
system with the interpolators
$\mathcal{O}^-_1,\mathcal{O}^-_2,\mathcal{O}^-_4,\mathcal{O}^-_5,\mathcal{O}^-_7,\mathcal{O}^-_8,\mathcal{O}^-_9$.
The energy levels are determined by correlated 
one-exponential fits to the eigenvalues  $\lambda_n(t)$
in the given fit range. 
We verified that two-exponential fits starting at smaller $t$
agree with results obtained from one-exponential fits. The errors are
determined by the single-elimination jackknife method.
For the values given in GeV we use the lattice spacing $a=0.1239$ fm (Sommer
parameter $r_0=0.48$fm).}\label{tab:results}
\end{table*}

Figure \ref{fig:comparison} (right) demonstrates the difference to the previous case with only
3-quark interpolators.
The lowest level now lies slightly below threshold, a feature typical for
attractive $s$-wave \cite{Lang:2012sv,Mohler:2012na} and a finite volume
artifact. This agrees with the
behavior discussed in Subsect. \ref{subject:levelsinterpretation}. The
next-higher two levels are now close to values lying approximately 130 MeV above
the physical resonance positions of $N^*(1535)$ and $N^*(1650)$,
similar to the situation for the nucleon. Comparison
with Fig.~\ref{fig:energylevels}, where a single elastic resonance parameterization
has been used, shows excellent agreement for the lowest two energy levels.

\begin{figure}[t]
\begin{center}
\includegraphics[width=0.48\textwidth,clip]{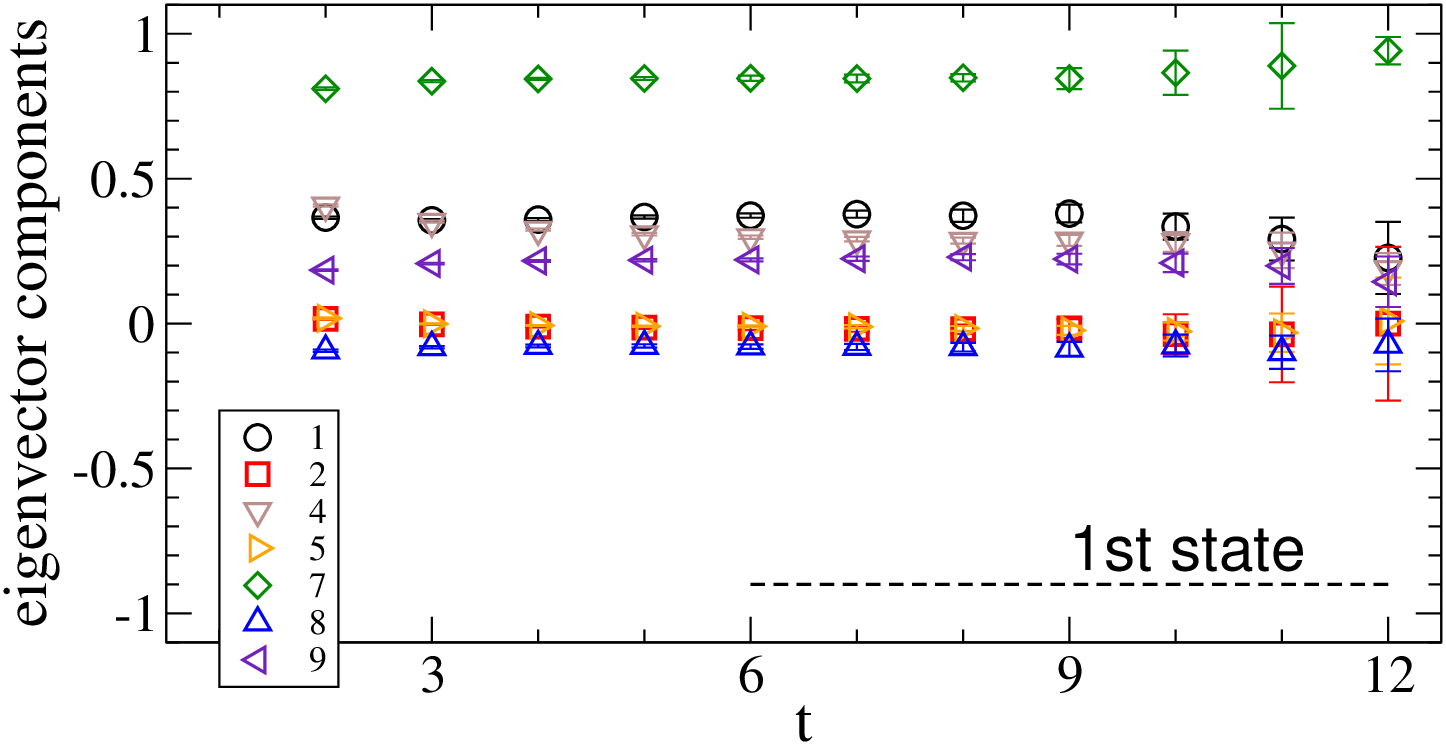}\\
\includegraphics[width=0.48\textwidth,clip]{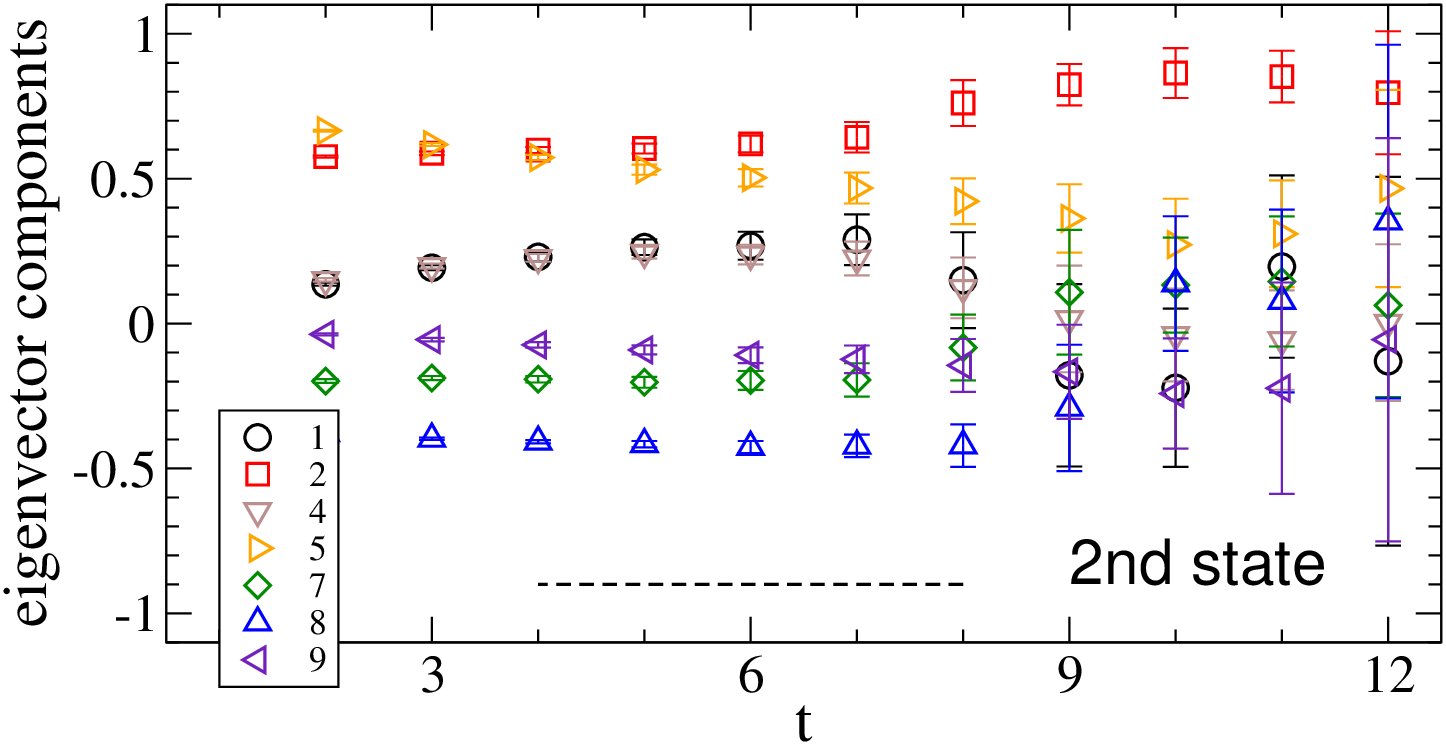}\\
\includegraphics[width=0.48\textwidth,clip]{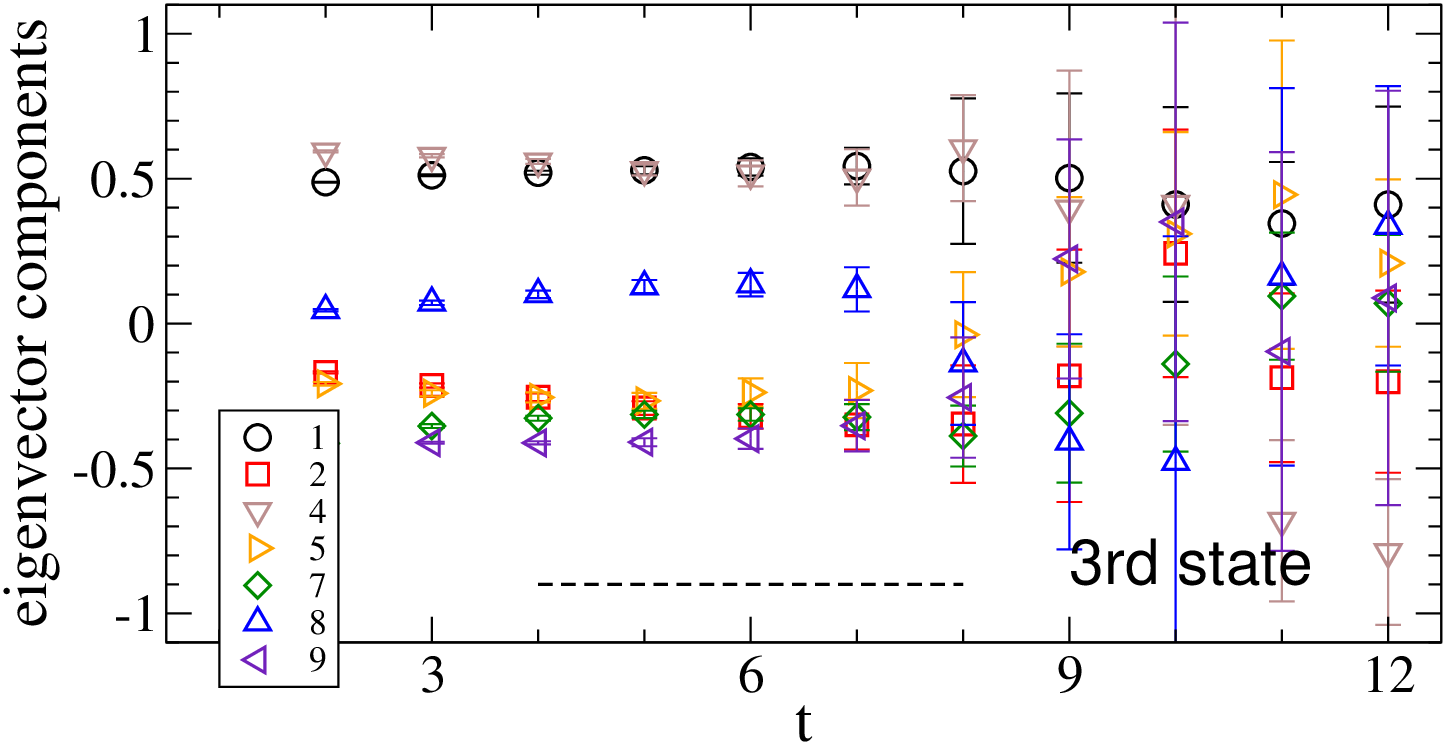}
\end{center}
\caption{The (normalized) eigenvector components for the lowest three
eigenstates observed; the $t$-range used for the exponential fit to the
eigenvalues is indicated by a broken line. In the legends the operator numbers
according  to \eq{eq:operators} are given.}\label{fig:eigenvectors}
\end{figure}

The eigenvectors are fingerprints of the states and one should have a stable
composition across the fit range in order to be sure to identify the same
eigenstate. Fig.~\ref{fig:eigenvectors} shows the eigenvector components of the
three lowest eigenstates. The eigenvectors have unit norm. The absolute
normalization of the 5-quark operators compared to the 3-quark ones is unclear. 
However, one finds that the $\mathcal{O}_{\Npi}$ contribution to the ground
state is significantly larger than to the higher levels. Interpolators of type
$N_-^{(1)}$ contribute importantly to the lowest eigenstate and dominate
the 3rd state, whereas the interpolators of type $N_-^{(2)}$ are more important
for the 2nd state.

In contrast, the effective energy levels of the pure 3-quark 
correlations system show more
fluctuation. Comparing with the full $\Npi$ system results one gets the
impression that the two lowest states of the 3-quark system interpolate
between the three lowest states  of the complete system.

The lowest energy level of the two particle system lies below threshold and the
corresponding value of $\rho_0$ may be related to the scattering length.  Table 
\ref{tab:results} gives also the values of $\rho_0$ from \eq{eq:rho0} due to the
L\"uscher analysis and the resulting values of the phase shift, assuming
elasticity. The second energy level lies close to the point where the  phase
shift crosses $\pi/2$ (this value is included within the error bars). 
This closeness is pure chance: for slightly larger lattices this would not have been the case
(cf., Fig.~\ref{fig:energylevels}).  As discussed,
the kinematical situation (pion mass and lattice size) allows the assumption to
be in the elastic domain and thus one is tempted to assume validity of
\eq{eq:BW}. The zero of the line connecting the values of $\rho_0$ at the two
lowest energy levels give the resonance position $a^2 s_R=1.114(135)$
corresponding to a  resonance mass $m_R=1.678(99)$ GeV. 
This is approximately 140
MeV above the physical value, but not surprising due to the unphysical pion and
nucleon masses, in fact, a similar shift as for the nucleon. Also note, that the
$\Npi$ system in Nature is already inelastic and the linearity assumption not
justified in that case.

The third eigenstate has a phase shift of $47^\circ$ ($\simeq 227^\circ$, since
the $\arctan$ is defined modulo $180^\circ$), indicating a resonance lying
closely above that energy value of 1.79 GeV -- again
assuming elastic scattering.

Due to the closeness of the  threshold to the resonance in our setting, as
compared to Nature, we cannot expect physical values for scattering length or
decay width. With \eq{eq:scattlength} we can estimate the scattering length from
the point close below threshold $s_{thr}$. We  find a value $a_0\simeq
5.3(\pm1.4)\;\textrm{GeV}^{-1}$ roughly four times larger than, e.g., the
leading order Chiral Perturbation Theory value $m_\pi/(4 \pi F_\pi^2)$
\cite{Weinberg:1966kf,Tomozawa:1966jm}.

\section{Summary}
\label{Sec_Conclusion}

We studied $\Npi$ scattering in the negative parity, isospin $\ot$ sector in an
ab initio lattice QCD calculation. The simulation parameters  are: two
dynamical, mass degenerate quarks, a pion mass of 266 MeV, a spatial lattice
size of $1.98$ fm, a volume $16^3\times 32$ in lattice units. We use 3-quark and
meson-baryon (5 quark) interpolators and analyze the $9\times 9$ correlation
matrix with help of the variational method. 

We find a significant difference to the results of simple 3-quark correlation
analyses. The overall behavior is resembling that found in meson-meson
scattering lattice studies for $s$-wave channels
\cite{Lang:2012sv,Mohler:2012na}. Due to the unphysical values of the pion mass,
the resonance position is higher than the experimentally established values. 

The main result of our study is that taking into account meson-baryon
interpolators indeed changes the obtained energy spectrum significantly. This is
a first step into that direction. Obviously this is an exploratory study and
systematic uncertainties stemming from the volume size, the lattice spacing and
the pion mass are not (yet) under control. More work (moving frames, different
volumes, further coupled channels) will fill the gap between
elastic and inelastic threshold and allow the comparison with experiment and
continuum models.


\acknowledgments
We are grateful to Meinulf G\"ockeler and Akaki Rusetsky for several helpful
discussions and suggestions. We would like to thank Georg Engel, Christof
Gattringer, Leonid Glozman, Daniel Mohler, Colin Morningstar and Sasa
Prelovsek  for  many discussions. Thanks also to Anna Hasenfratz for
providing the dynamical configurations and to Daniel Mohler and Sasa
Prelovsek  for allowing us to use the perambulators derived in another
project. The calculations were performed on local clusters at UNI-IT at the
University of Graz.  V.V.~ has been supported by the Austrian Science Fund
(FWF) under Grant DK W1203-N16.

\begin{appendix}
\section{Wick contractions}\label{app:contractions}

Notation for the perambulators used in this section: $\tau\left(t,t',a,a',\alpha
,\alpha '\right)$ denotes the perambulator  $\tau_{\alpha\alpha'}(a,t;a',t')$
from \eq{eq:deftau}, i.e., from source at $t'$ (source vector $a'$, Dirac index
$\alpha'$) to the sink at $t$ (source vector $a$, Dirac index $\alpha$).  

Each source/sink nucleon contributes a factor of the form
$\widehat\phi_N(a,b,c)$, which  is constructed from the  Laplacian eigenvectors.
For a given time slice we have
\be
\widehat\phi_N^{snk}(a,b,c)=\sum_{\vec x,i,j,k} 
\epsilon_{ijk}v_a^i(\vec x)v_b^j(\vec x)v_c^k(\vec x)\;,
\ee
where $\epsilon$ denotes the Levi-Civita symbol, $v$ are the Laplacian
eigenvectors, and  the sum runs over all sites of the time slice and over the 
color indices $i,j,k$. The corresponding factor for the pion
$\widehat\phi_\pi(a,b)$ on a given time slice reads
\be
\widehat\phi_\pi^{snk}(a,b)=\sum_{\vec x,i,j} 
\delta_{ij} v_a^{i*}(\vec x)v_b^j(\vec x)\;.
\ee
By permuting and renaming the Dirac indices $\alpha,\beta, \gamma,\ldots$ and the eigenvector
indices  $a,b,c,\ldots$ we group the different  contraction such that they have
a common prefactor. There also the  gamma matrices of the nucleon and pion and
the parity projection operators $P^\pm$ are located. 
 
\subsection{$N\to N$}
This entry has the form
\be
\Gamma^{A\dagger}_{\alpha'\mu}P^\pm_{\mu\nu}\Gamma^{A}_{\nu\alpha} 
\Gamma^B_{\beta\gamma}\Gamma^{B\dagger}_{\gamma'\beta'} 
\widehat\phi_N^{snk}(a,b,c)
\widehat\phi_N^{src}(a',b',c')\;\sum_{i=1}^2 A_i\FC
\ee
where summation over index pairs is implied.

\begin{figure}[h]
\begin{center}
\includegraphics[width=0.35\textwidth,clip]{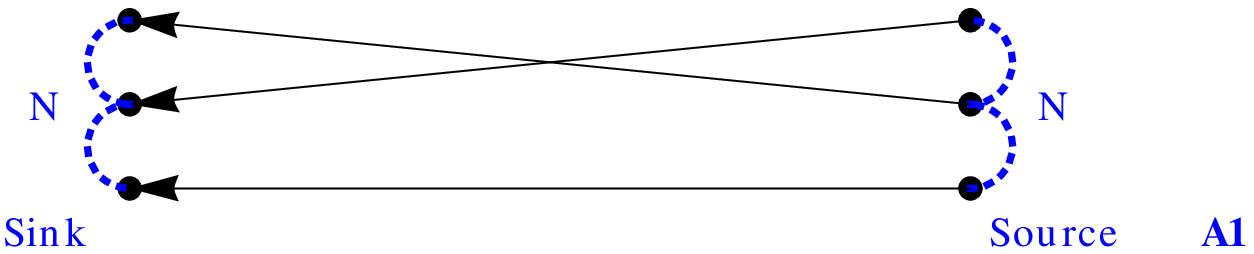}\\
\includegraphics[width=0.35\textwidth,clip]{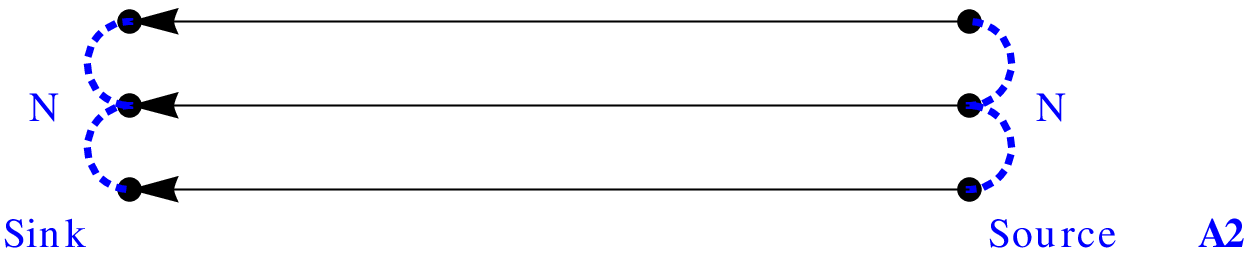}
\end{center}
\caption{Terms $A_1$ and $A_2$ contributing to $N\to N$.}
\end{figure}

\begin{align}
A_{1} &= \tau\left(t,t',c,c',\gamma ,\gamma '\right)         \,\tau\left(t,t',a,b',\alpha ,\beta '\right)   \nonumber\\
  &  \qquad\,\tau\left(t,t',b,a',\beta ,\alpha'\right)\nonumber\\
A_{2} &= -\tau\left(t,t',a,a',\alpha ,\alpha '\right)        \,\tau\left(t,t',b,b',\beta ,\beta '\right)    \nonumber\\
  &  \qquad\,\tau\left(t,t',c,c',\gamma ,\gamma'\right)
\end{align}

\subsection{$N\to \Npi$}
This matrix element has 4 terms contributing:
\begin{align}
\frac{1}{\sqrt{2}}
&\Gamma^{A\dagger}_{\alpha'\mu}P^\pm_{\mu\nu}\Gamma^{A}_{\nu\alpha} 
\Gamma^B_{\beta\gamma}\Gamma^{B\dagger}_{\gamma'\beta'} 
\Gamma^{\pi}_{\delta\epsilon}\nonumber\\
&\widehat\phi_N^{snk}(a,b,c)
\widehat\phi_\pi^{snk}(e, g)
\widehat\phi_N^{src}(a',b',c')\;\sum_{i=1}^4 B_i \FC
\end{align}
where summation over index pairs is implied.

\begin{figure}[h]
\begin{center}
\includegraphics[width=0.35\textwidth,clip]{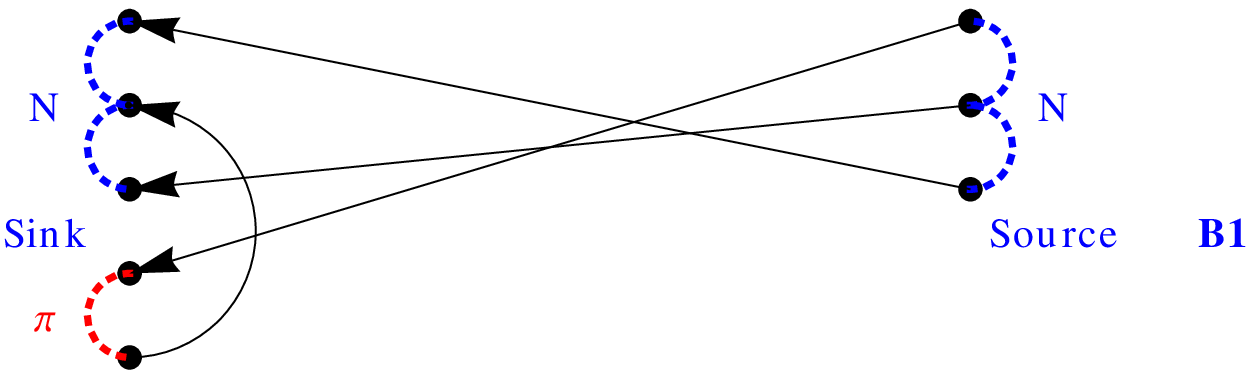}\\
\includegraphics[width=0.35\textwidth,clip]{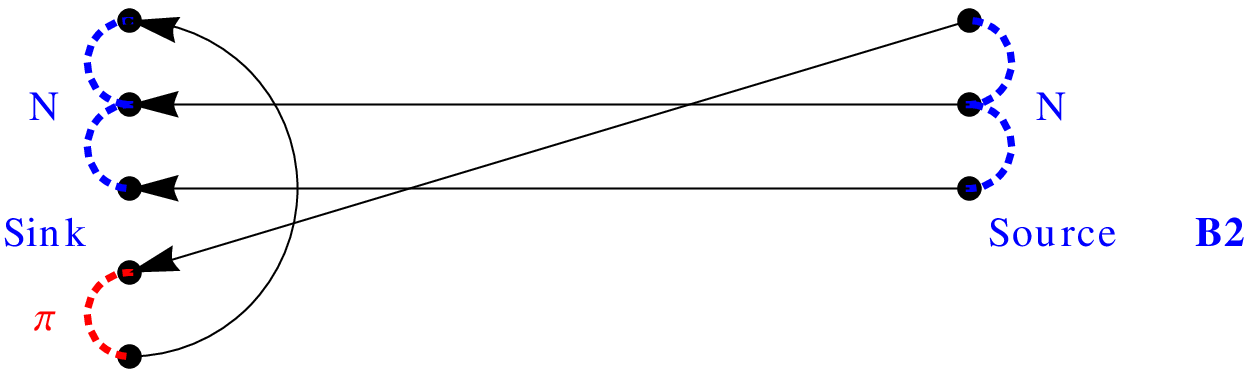}\\
\includegraphics[width=0.35\textwidth,clip]{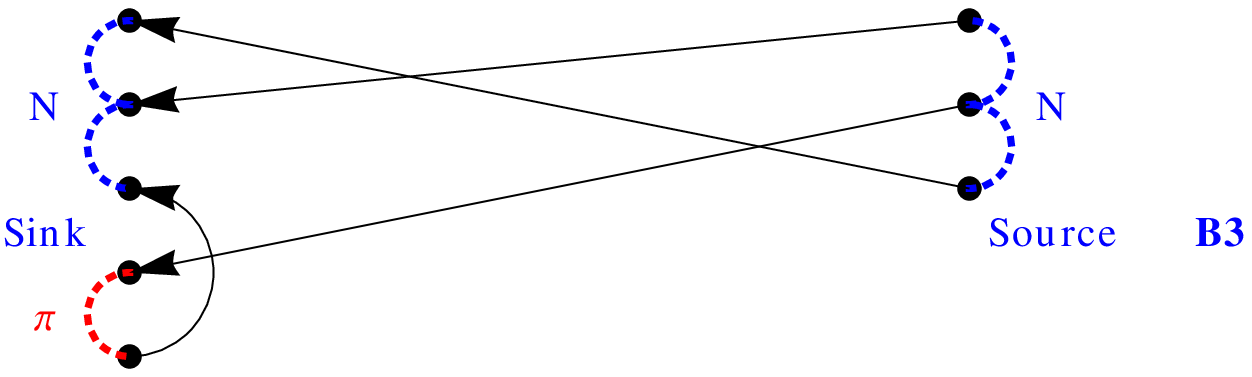}\\
\includegraphics[width=0.35\textwidth,clip]{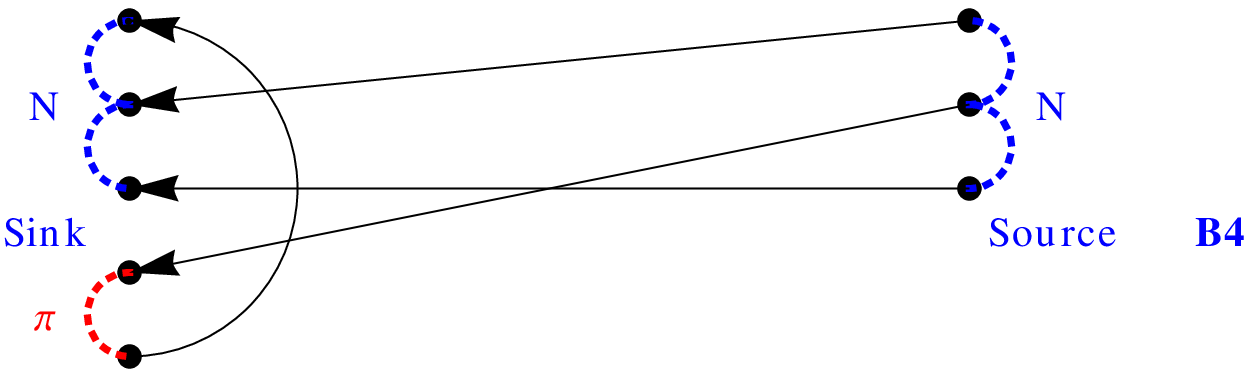}
\end{center}
\caption{Terms $B_1-B_4$ contributing to $N\to \Npi$.}
\end{figure}

\begin{align}
B_{1} &= 3 \,\tau\left(t,t,b,e,\beta ,\epsilon\right)           \,\tau\left(t,t',a,c',\alpha ,\gamma '\right)\nonumber\\
  &  \qquad\,\tau\left(t,t',g,a',\delta ,\alpha '\right)        \,\tau\left(t,t',c,b',\gamma ,\beta '\right)\nonumber\\
B_{2} &= -3\,\tau\left(t,t',b,b',\beta ,\beta '\right)          \,\tau\left(t,t',c,c',\gamma ,\gamma '\right)\nonumber\\
  &  \qquad\,\tau\left(t,t,a,e,\alpha ,\epsilon\right)          \,\tau\left(t,t',g,a',\delta ,\alpha '\right)\nonumber\\
B_{3} &= -3\,\tau\left(t,t,c,e,\gamma ,\epsilon\right)          \,\tau\left(t,t',b,a',\beta ,\alpha '\right)   \nonumber\\
  &  \qquad\,\tau\left(t,t',a,c',\alpha ,\gamma '\right)        \,\tau\left(t,t',g,b',\delta ,\beta '\right)\nonumber\\
B_{4} &= 3 \,\tau\left(t,t',c,c',\gamma ,\gamma '\right)        \,\tau\left(t,t,a,e,\alpha ,\epsilon\right)   \nonumber\\
  &  \qquad \,\tau\left(t,t',b,a',\beta ,\alpha '\right)        \,\tau\left(t,t',g,b',\delta ,\beta '\right)
\end{align}

\subsection{$\Npi\to N$}
This matrix element has 4 terms contributing:
\begin{align}
\frac{1}{\sqrt{2}}
&\Gamma^{A\dagger}_{\alpha'\mu}P^\pm_{\mu\nu}\Gamma^{A}_{\nu\alpha} 
\Gamma^B_{\beta\gamma}\Gamma^{B\dagger}_{\gamma'\beta'} 
\Gamma^{\pi\dagger}_{\epsilon'\delta'}\nonumber\\
&\widehat\phi_N^{snk}(a,b,c)
\widehat\phi_N^{src}(a',b',c')
\widehat\phi_\pi^{src}(g',e')\;\sum_{i=1}^4 C_i \FC
\end{align}
where summation over index pairs is implied.

\begin{figure}[h]
\begin{center}
\includegraphics[width=0.35\textwidth,clip]{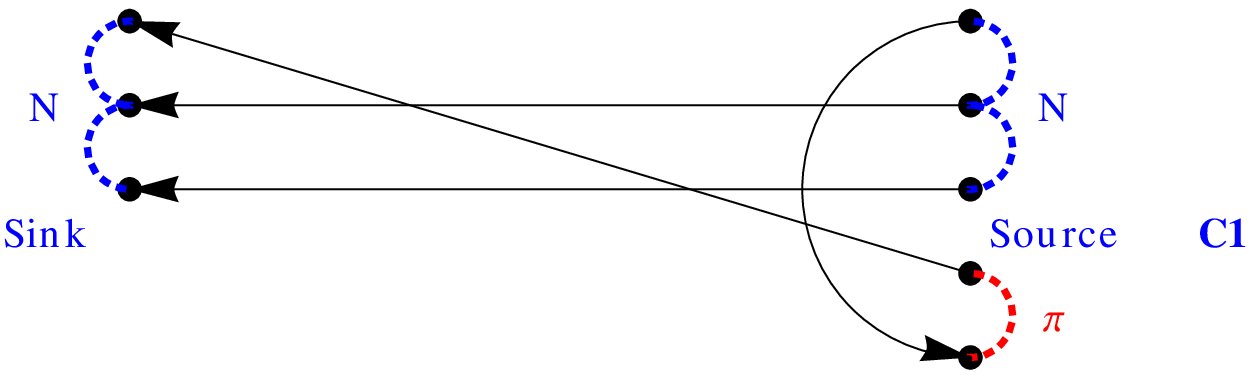}\\
\includegraphics[width=0.35\textwidth,clip]{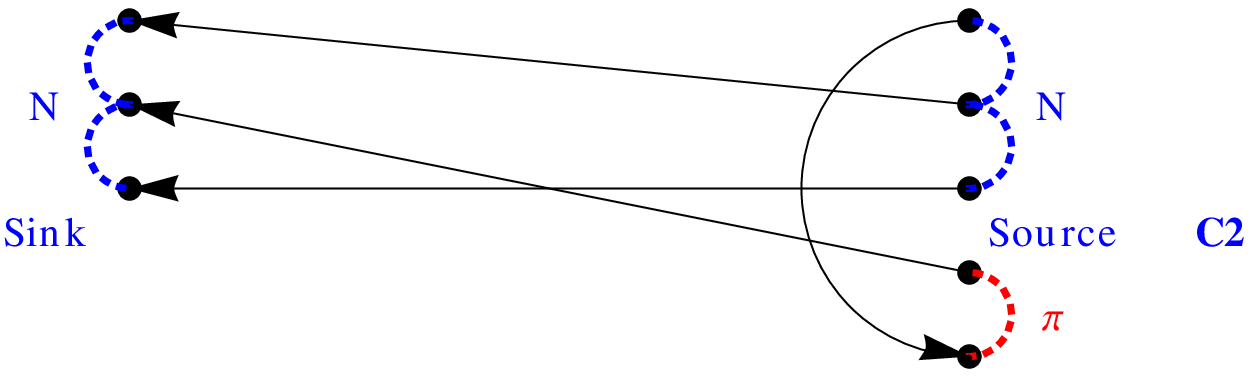}\\
\includegraphics[width=0.35\textwidth,clip]{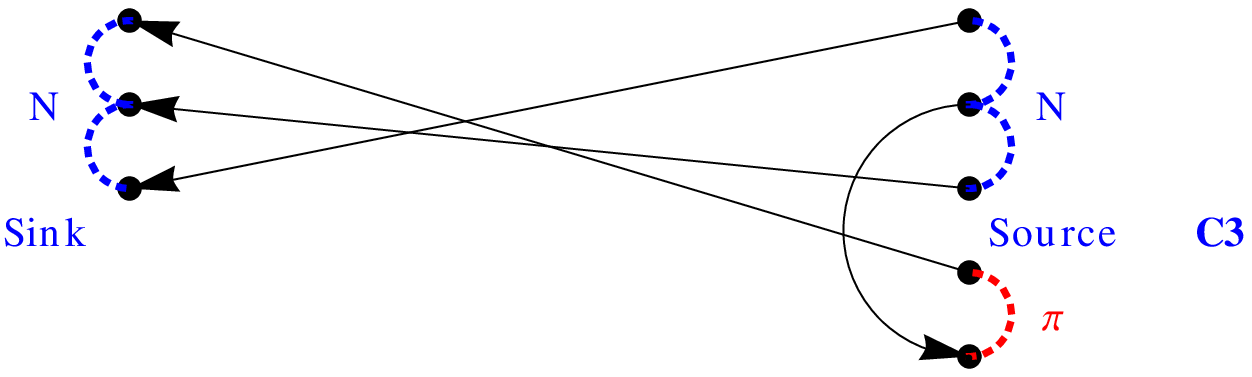}\\
\includegraphics[width=0.35\textwidth,clip]{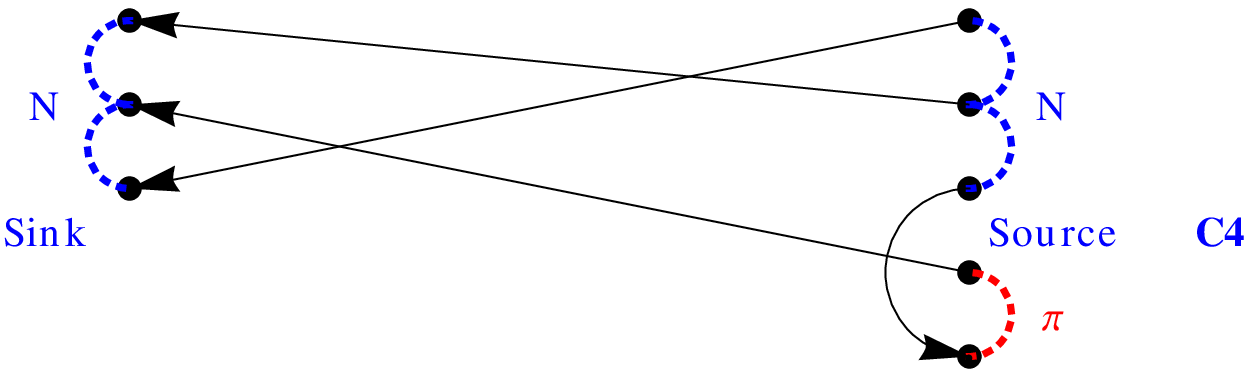}
\end{center}
\caption{Terms $C_1-C_4$ contributing to $\Npi\to N$.}
\end{figure}

\begin{align}
C_{1} &= 3 \,\tau\left(t,t',b,b',\beta ,\beta '\right)          \,\tau\left(t,t',c,c',\gamma ,\gamma '\right) \nonumber\\
  &  \qquad\,\tau\left(t',t',e',a',\epsilon ',\alpha '\right)   \,\tau\left(t,t',a,g',\alpha ,\delta '\right)\nonumber\\
C_{2} &= -3\,\tau\left(t,t',c,c',\gamma ,\gamma '\right)        \,\tau\left(t,t',a,b',\alpha ,\beta '\right)  \nonumber\\
  &  \qquad\,\tau\left(t',t',e',a',\epsilon ',\alpha '\right)   \,\tau\left(t,t',b,g',\beta ,\delta '\right)\nonumber\\
C_{3} &= -3\,\tau\left(t,t',c,a',\gamma ,\alpha '\right)        \,\tau\left(t,t',a,g',\alpha ,\delta '\right) \nonumber\\
  &  \qquad\,\tau\left(t,t',b,c',\beta ,\gamma '\right)         \,\tau\left(t',t',e',b',\epsilon ',\beta '\right)\nonumber\\
C_{4} &= 3 \,\tau\left(t,t',a,b',\alpha ,\beta '\right)         \,\tau\left(t,t',c,a',\gamma ,\alpha '\right) \nonumber\\
  &  \qquad\,\tau\left(t,t',b,g',\beta ,\delta '\right)         \,\tau\left(t',t',e',c',\epsilon ',\gamma '\right)
\end{align}

\subsection{$\Npi\to \Npi$}
Here 19 terms contribute:
\begin{align}
\frac{1}{2}
&\Gamma^{A\dagger}_{\alpha'\mu}P^\pm_{\mu\nu}\Gamma^{A}_{\nu\alpha} 
\Gamma^B_{\beta\gamma}\Gamma^{B\dagger}_{\gamma'\beta'} 
\Gamma^\pi_{\delta\epsilon}
\Gamma^{\pi\dagger}_{\epsilon'\delta'}\nonumber\\
&\widehat\phi_N^{snk}(a,b,c)
\widehat\phi_\pi^{snk}(e,g)
\widehat\phi_N^{src}(a',b',c')
\widehat\phi_\pi^{src}(g',e')\;\sum_{i=1}^{19} D_i \FC
\end{align}
where summation over index pairs is implied.

\begin{figure}[h]
\begin{center}
\includegraphics[width=0.35\textwidth,clip]{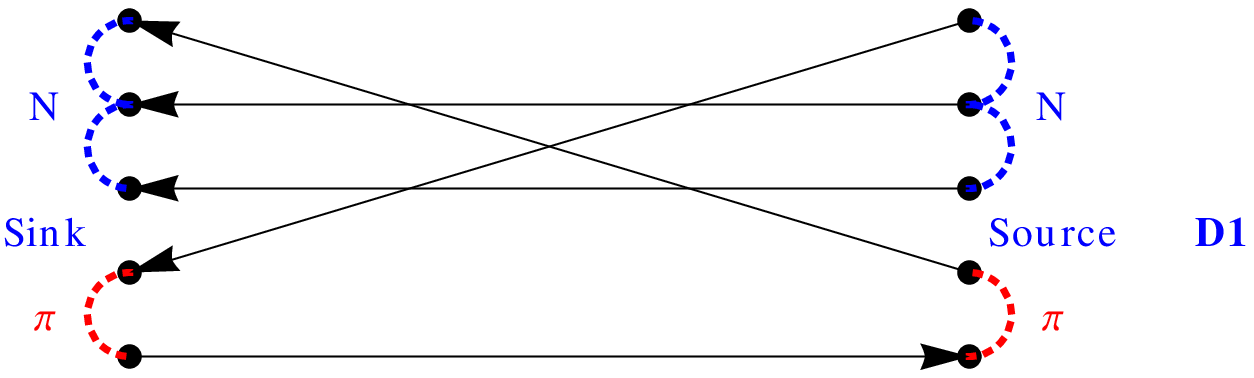}\\
\includegraphics[width=0.35\textwidth,clip]{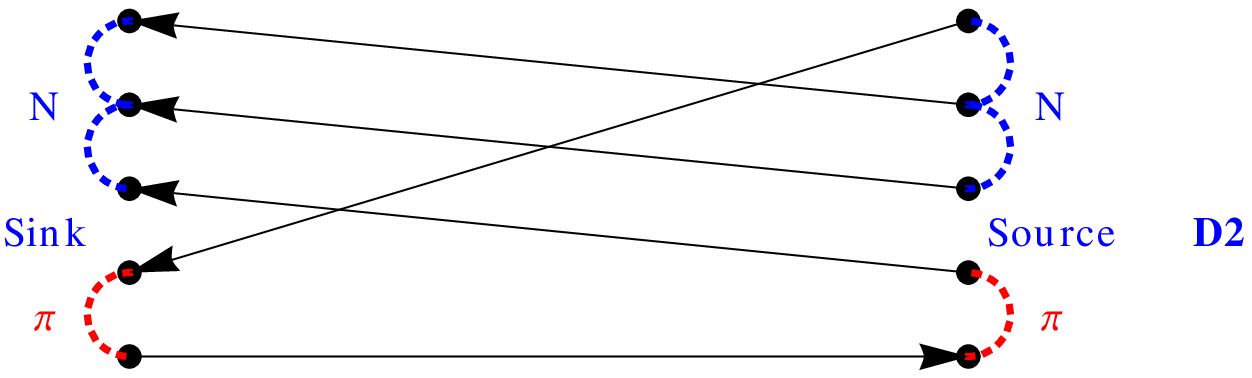}\\
\includegraphics[width=0.35\textwidth,clip]{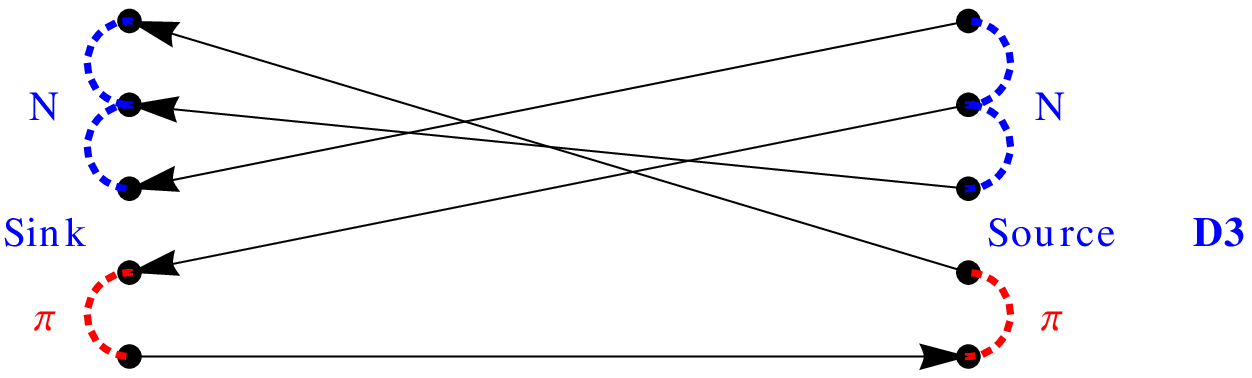}\\
\includegraphics[width=0.35\textwidth,clip]{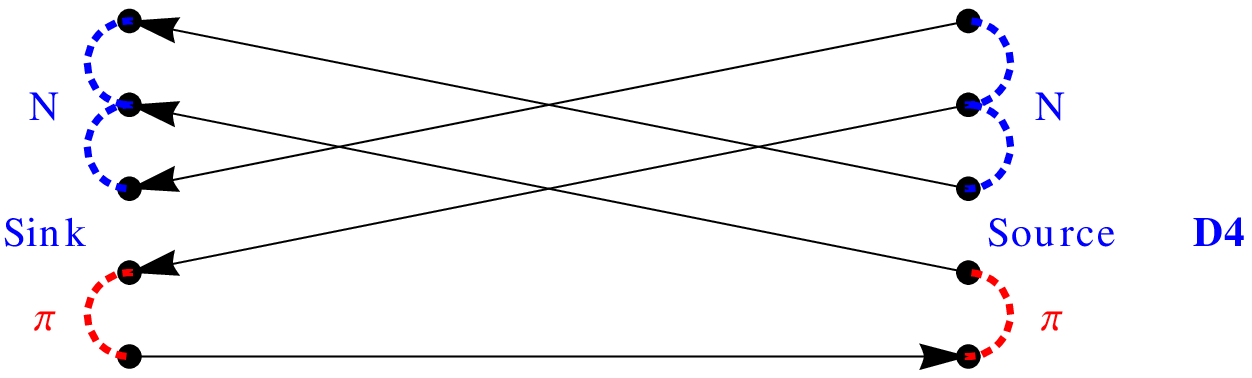}\\
\includegraphics[width=0.35\textwidth,clip]{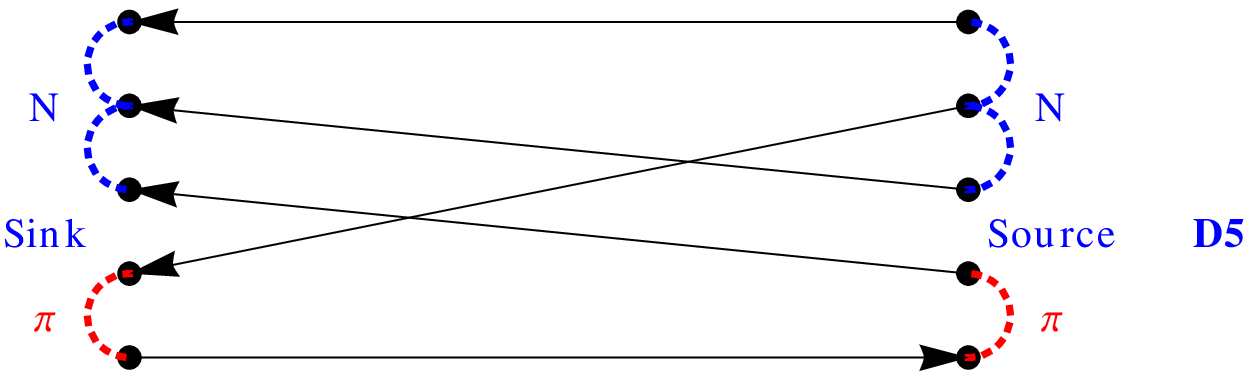}
\end{center}
\caption{Terms $D_1-D_5$ contributing to $\Npi\to \Npi$.}
\end{figure}

\begin{align}
D_{1} &= 3 \,\tau\left(t,t',b,b',\beta ,\beta '\right)       \,\tau\left(t,t',c,c',\gamma ,\gamma '\right)      \nonumber\\
  &  \qquad\,\tau\left(t',t,e',e,\epsilon ',\epsilon \right)      \tau\left(t,t',a,g',\alpha ,\delta '\right)      \nonumber\\
  &  \qquad\,\tau\left(t,t',g,a',\delta ,\alpha '\right)\nonumber\\
D_{2} &= -3 \,\tau\left(t',t,e',e,\epsilon ',\epsilon \right)\,\tau\left(t,t',a,b',\alpha ,\beta '\right)       \nonumber\\
  &  \qquad\,\tau\left(t,t',g,a',\delta ,\alpha '\right)          \tau\left(t,t',b,c',\beta ,\gamma '\right)       \nonumber\\
  &  \qquad\,\tau\left(t,t',c,g',\gamma ,\delta '\right)\nonumber\\
D_{3} &= -3 \,\tau\left(t',t,e',e,\epsilon ',\epsilon \right)\,\tau\left(t,t',c,a',\gamma ,\alpha '\right)      \nonumber\\
  &  \qquad\,\tau\left(t,t',a,g',\alpha ,\delta '\right)          \tau\left(t,t',b,c',\beta ,\gamma '\right)       \nonumber\\
  &  \qquad\,\tau\left(t,t',g,b',\delta ,\beta '\right)\nonumber\\
D_{4} &= 9 \,\tau\left(t',t,e',e,\epsilon ',\epsilon \right) \,\tau\left(t,t',a,c',\alpha ,\gamma '\right)      \nonumber\\
  &  \qquad\,\tau\left(t,t',c,a',\gamma ,\alpha '\right)          \tau\left(t,t',b,g',\beta ,\delta '\right)       \nonumber\\
  &  \qquad\,\tau\left(t,t',g,b',\delta ,\beta '\right)\nonumber\\
D_{5} &= -6 \,\tau\left(t,t',a,a',\alpha ,\alpha '\right)    \,\tau\left(t',t,e',e,\epsilon ',\epsilon \right)  \nonumber\\
  &  \qquad\,\tau\left(t,t',b,c',\beta ,\gamma '\right)           \tau\left(t,t',g,b',\delta ,\beta '\right)       \nonumber\\
  &  \qquad\,\tau\left(t,t',c,g',\gamma ,\delta '\right)
\end{align}

\begin{figure}[h]
\begin{center}
\includegraphics[width=0.35\textwidth,clip]{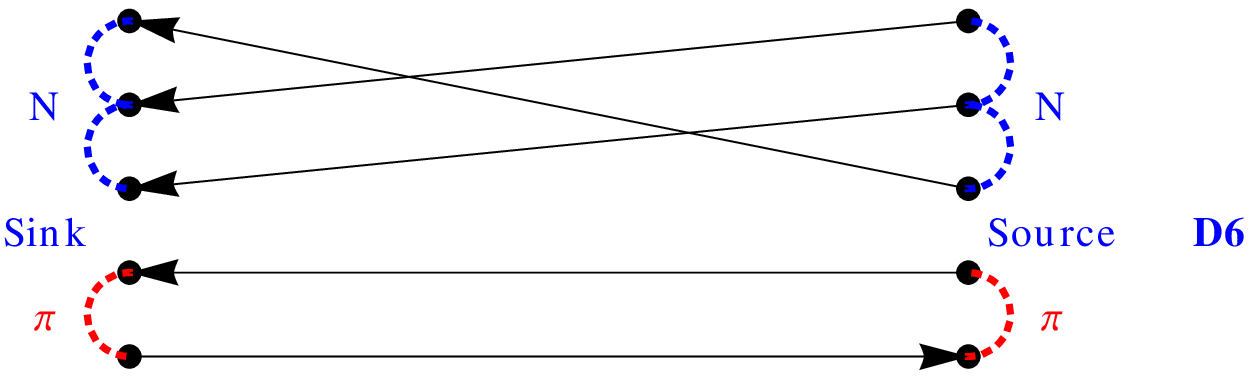}\\
\includegraphics[width=0.35\textwidth,clip]{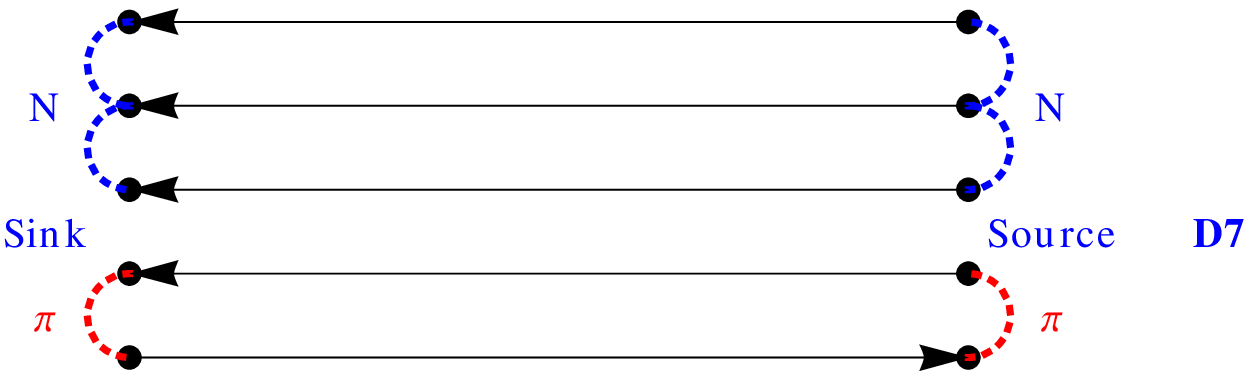}\\
\includegraphics[width=0.35\textwidth,clip]{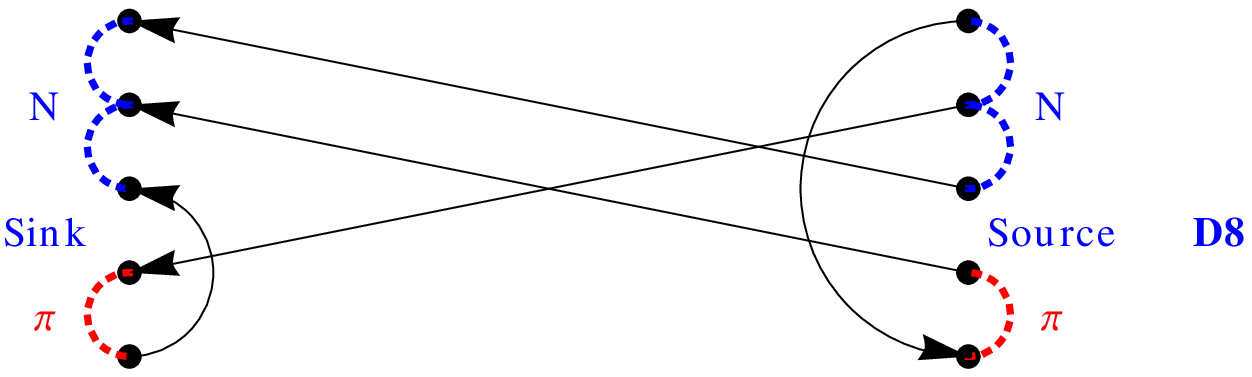}\\
\includegraphics[width=0.35\textwidth,clip]{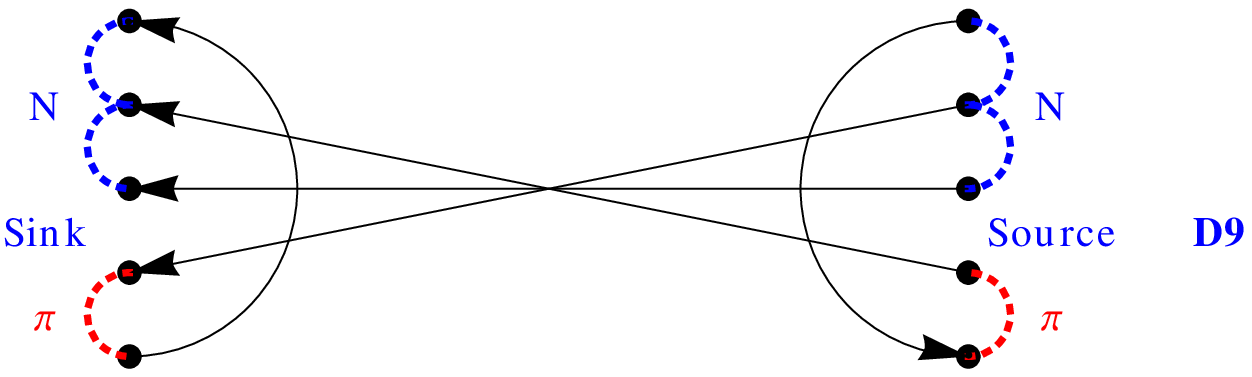}\\
\includegraphics[width=0.35\textwidth,clip]{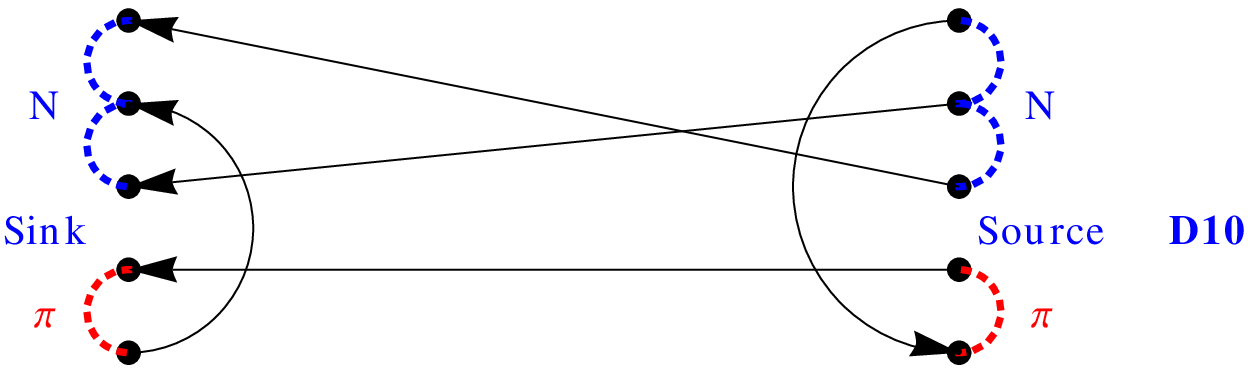}
\end{center}
\caption{Terms $D_6-D_{10}$ contributing to $\Npi\to \Npi$}
\end{figure}

\begin{align}
D_{6} &= -6 \,\tau\left(t',t,e',e,\epsilon ',\epsilon \right)\,\tau\left(t,t',g,g',\delta ,\delta '\right)      \nonumber\\
  &  \qquad\,\tau\left(t,t',b,a',\beta ,\alpha '\right)           \tau\left(t,t',a,c',\alpha ,\gamma '\right)      \nonumber\\
  &  \qquad\,\tau\left(t,t',c,b',\gamma ,\beta '\right)\nonumber\\
D_{7} &= 6 \,\tau\left(t,t',a,a',\alpha ,\alpha '\right)     \,\tau\left(t,t',b,b',\beta ,\beta '\right)        \nonumber\\
  &  \qquad\,\tau\left(t,t',c,c',\gamma ,\gamma '\right)          \tau\left(t',t,e',e,\epsilon ',\epsilon \right)  \nonumber\\
  &  \qquad\,\tau\left(t,t',g,g',\delta ,\delta '\right)\nonumber\\
D_{8} &= -9 \,\tau\left(t,t,c,e,\gamma ,\epsilon\right)      \,\tau\left(t,t',a,c',\alpha ,\gamma '\right)      \nonumber\\
  &  \qquad\,\tau\left(t',t',e',a',\epsilon ',\alpha '\right)     \tau\left(t,t',b,g',\beta ,\delta '\right)       \nonumber\\
  &  \qquad\,\tau\left(t,t',g,b',\delta ,\beta '\right)\nonumber\\
D_{9} &= 9 \,\tau\left(t,t',c,c',\gamma ,\gamma '\right)     \,\tau\left(t,t,a,e,\alpha ,\epsilon\right)        \nonumber\\
  &  \qquad\,\tau\left(t',t',e',a',\epsilon ',\alpha '\right)     \tau\left(t,t',b,g',\beta ,\delta '\right)       \nonumber\\
  &  \qquad\,\tau\left(t,t',g,b',\delta ,\beta '\right)\nonumber\\
D_{10} &= 9 \,\tau\left(t,t',g,g',\delta ,\delta '\right)    \,\tau\left(t,t,b,e,\beta ,\epsilon\right)         \nonumber\\
  &  \qquad\,\tau\left(t,t',a,c',\alpha ,\gamma '\right)          \tau\left(t',t',e',a',\epsilon ',\alpha '\right) \nonumber\\
  &  \qquad\,\tau\left(t,t',c,b',\gamma ,\beta '\right)
\end{align}

\begin{figure}[h]
\begin{center}
\includegraphics[width=0.35\textwidth,clip]{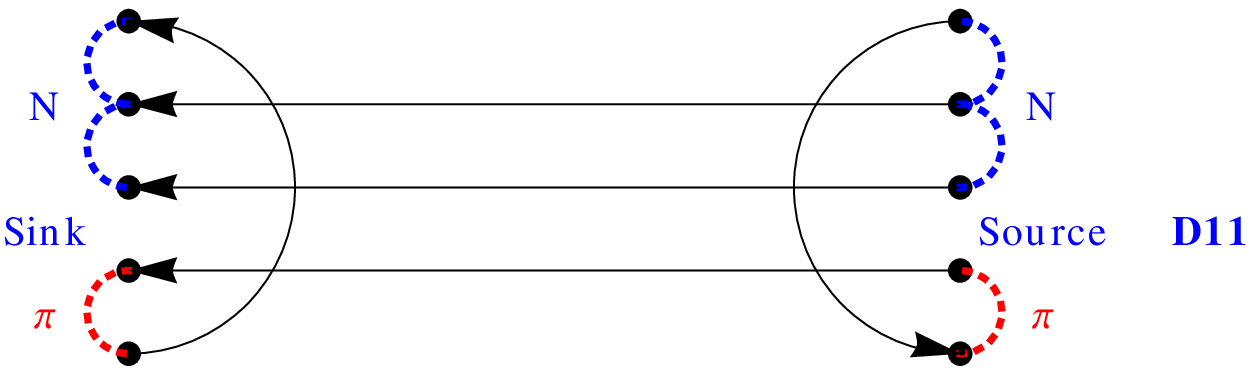}\\
\includegraphics[width=0.35\textwidth,clip]{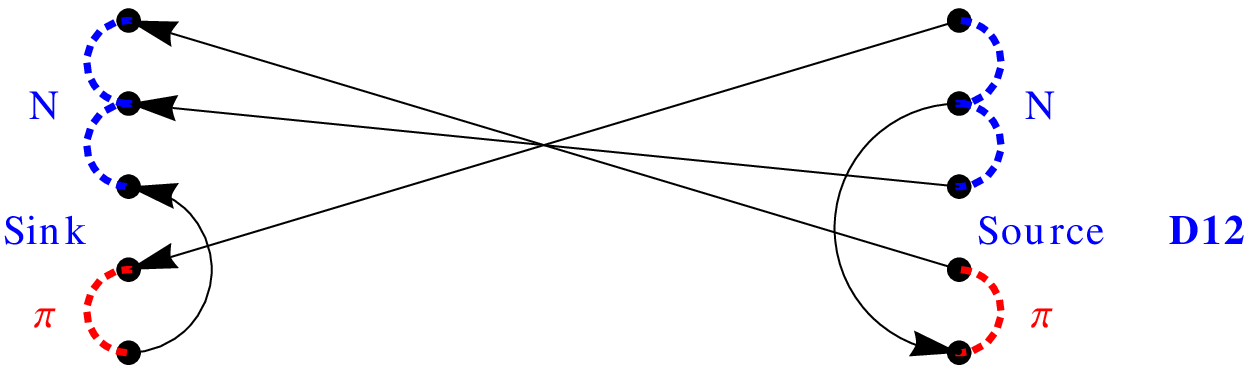}\\
\includegraphics[width=0.35\textwidth,clip]{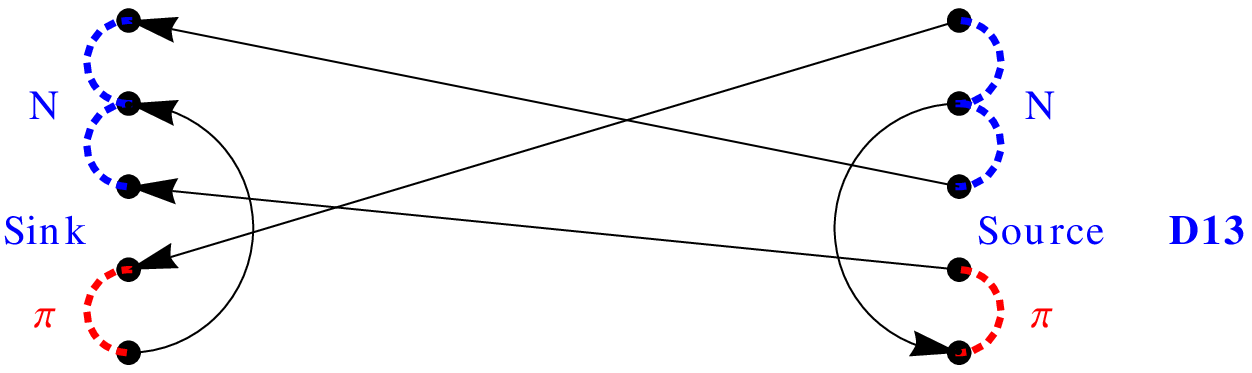}\\
\includegraphics[width=0.35\textwidth,clip]{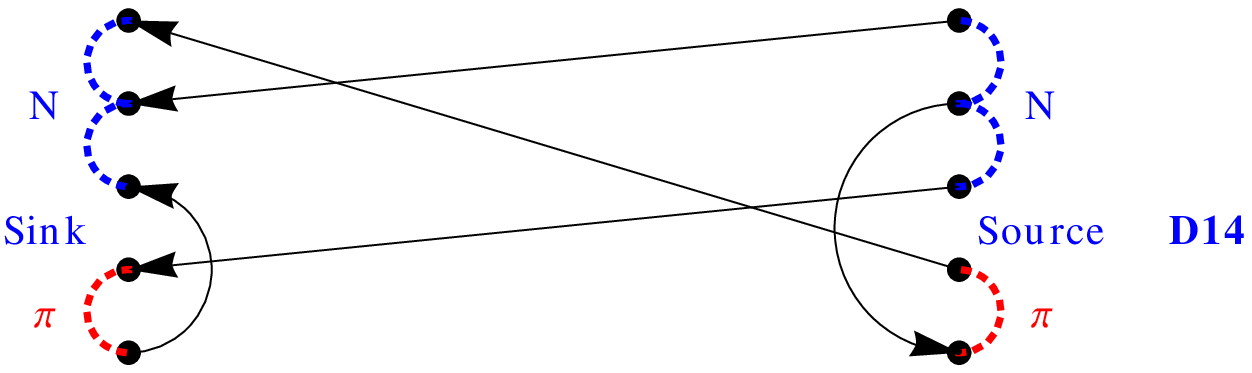}\\
\includegraphics[width=0.35\textwidth,clip]{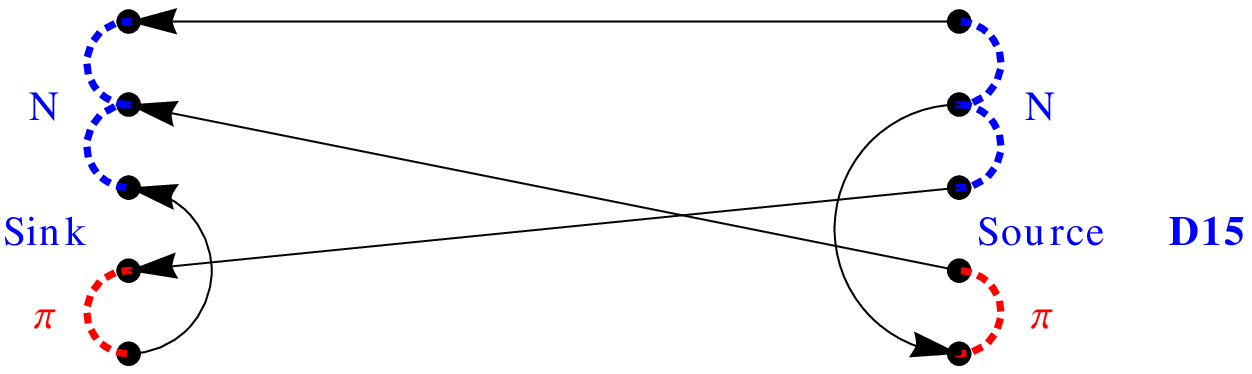}
\end{center}
\caption{Terms $D_{11}-D_{15}$ contributing to $\Npi\to \Npi$}
\end{figure}

\begin{align}
D_{11} &= -9 \,\tau\left(t,t',b,b',\beta ,\beta '\right)     \,\tau\left(t,t',c,c',\gamma ,\gamma '\right)      \nonumber\\
  &  \qquad\,\tau\left(t,t',g,g',\delta ,\delta '\right)          \tau\left(t,t,a,e,\alpha ,\epsilon\right)         \nonumber\\
  &  \qquad\,\tau\left(t',t',e',a',\epsilon ',\alpha '\right)\nonumber\\
D_{12} &= -3 \,\tau\left(t,t,c,e,\gamma ,\epsilon\right)     \,\tau\left(t,t',a,g',\alpha ,\delta '\right)      \nonumber\\
  &  \qquad\,\tau\left(t,t',g,a',\delta ,\alpha '\right)          \tau\left(t,t',b,c',\beta ,\gamma '\right)       \nonumber\\
  &  \qquad\,\tau\left(t',t',e',b',\epsilon ',\beta '\right)\nonumber\\
D_{13} &= 3 \,\tau\left(t,t,b,e,\beta ,\epsilon\right)       \,\tau\left(t,t',a,c',\alpha ,\gamma '\right)      \nonumber\\
  &  \qquad\,\tau\left(t,t',g,a',\delta ,\alpha '\right)          \tau\left(t',t',e',b',\epsilon ',\beta '\right)  \nonumber\\
  &  \qquad\,\tau\left(t,t',c,g',\gamma ,\delta '\right)\nonumber\\
D_{14} &= 3 \,\tau\left(t,t,c,e,\gamma ,\epsilon\right)      \,\tau\left(t,t',b,a',\beta ,\alpha '\right)       \nonumber\\
  &  \qquad\,\tau\left(t,t',a,g',\alpha ,\delta '\right)          \tau\left(t',t',e',b',\epsilon ',\beta '\right)  \nonumber\\
  &  \qquad\,\tau\left(t,t',g,c',\delta ,\gamma '\right)\nonumber\\
D_{15} &= 6 \,\tau\left(t,t',a,a',\alpha ,\alpha '\right)    \,\tau\left(t,t,c,e,\gamma ,\epsilon\right)        \nonumber\\
  &  \qquad\,\tau\left(t',t',e',b',\epsilon ',\beta '\right)      \tau\left(t,t',b,g',\beta ,\delta '\right)       \nonumber\\
  &  \qquad\,\tau\left(t,t',g,c',\delta ,\gamma '\right)
\end{align}

\begin{figure}[h]
\begin{center}
\includegraphics[width=0.35\textwidth,clip]{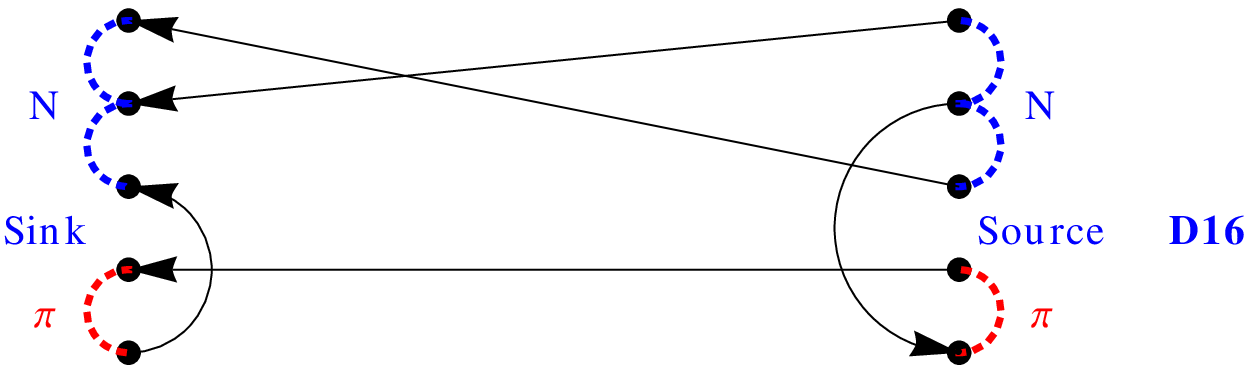}\\
\includegraphics[width=0.35\textwidth,clip]{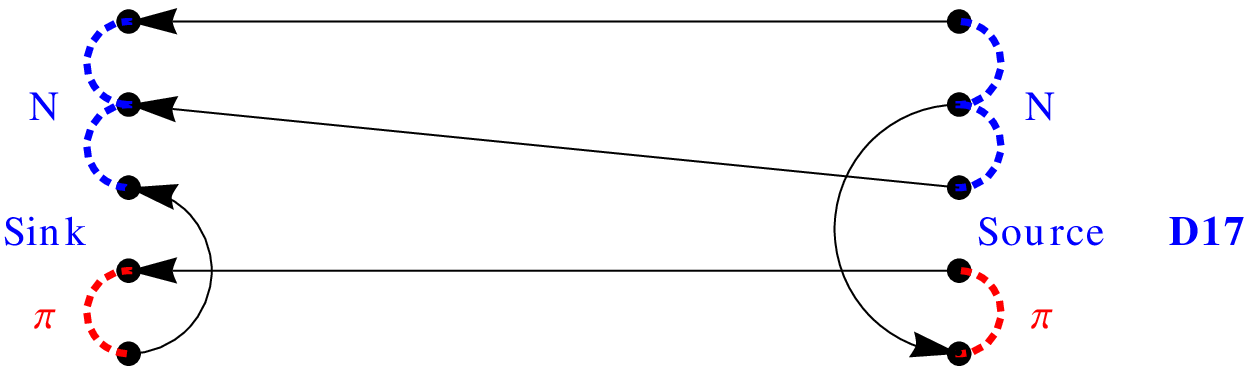}\\
\includegraphics[width=0.35\textwidth,clip]{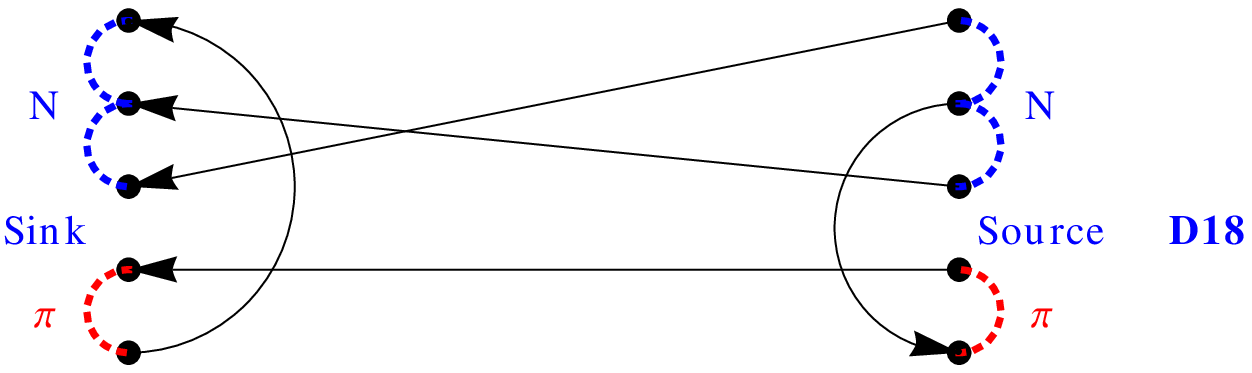}\\
\includegraphics[width=0.35\textwidth,clip]{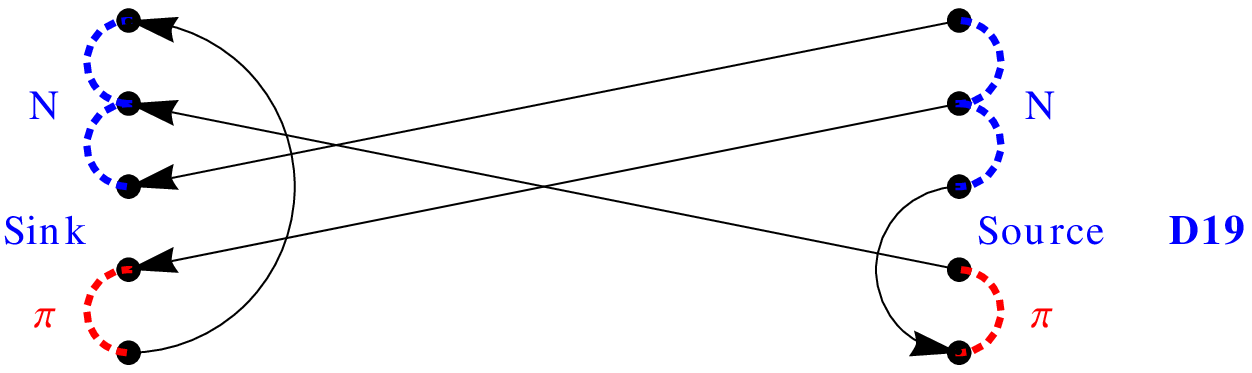}
\end{center}
\caption{Terms $D_{16}-D_{19}$ contributing to $\Npi\to \Npi$}
\end{figure}

\begin{align}
D_{16} &= -3 \,\tau\left(t,t',g,g',\delta ,\delta '\right)   \,\tau\left(t,t,c,e,\gamma ,\epsilon\right)        \nonumber\\
  &  \qquad\,\tau\left(t,t',b,a',\beta ,\alpha '\right)           \tau\left(t,t',a,c',\alpha ,\gamma '\right)      \nonumber\\
  &  \qquad\,\tau\left(t',t',e',b',\epsilon ',\beta '\right)\nonumber\\
D_{17} &= -6 \,\tau\left(t,t',a,a',\alpha ,\alpha '\right)   \,\tau\left(t,t',g,g',\delta ,\delta '\right)      \nonumber\\
  &  \qquad\,\tau\left(t,t,c,e,\gamma ,\epsilon\right)             \tau\left(t,t',b,c',\beta ,\gamma '\right)       \nonumber\\
  &  \qquad\,\tau\left(t',t',e',b',\epsilon ',\beta '\right)\nonumber\\
D_{18} &= 9 \,\tau\left(t,t',g,g',\delta ,\delta '\right)    \,\tau\left(t,t,a,e,\alpha ,\epsilon\right)        \nonumber\\
  &  \qquad\,\tau\left(t,t',c,a',\gamma ,\alpha '\right)          \tau\left(t,t',b,c',\beta ,\gamma '\right)       \nonumber\\
  &  \qquad\,\tau\left(t',t',e',b',\epsilon ',\beta '\right)\nonumber\\
D_{19} &= -9 \,\tau\left(t,t,a,e,\alpha ,\epsilon\right)     \,\tau\left(t,t',c,a',\gamma ,\alpha '\right)      \nonumber\\
  &  \qquad\,\tau\left(t,t',b,g',\beta ,\delta '\right)           \tau\left(t,t',g,b',\delta ,\beta '\right)       \nonumber\\
  &  \qquad\,\tau\left(t',t',e',c',\epsilon ',\gamma '\right)
\end{align}

\end{appendix}


\begin{thebibliography}{10}

\bibitem{Cohen:2009zk}
S.~Cohen {\em et~al.},
\newblock PoS {\bf LAT2009}, 112 (2009), [arXiv:0911.3373].

\bibitem{Mahbub:2009cf}
M.~S. Mahbub {\em et~al.},
\newblock PoS {\bf LAT2009}, 118 (2009), [arXiv:0910.2789].

\bibitem{Dudek:2010wm}
J.~J. Dudek, R.~G. Edwards, M.~J. Peardon, D.~G. Richards and C.~E. Thomas,
\newblock Phys. Rev. D {\bf 82}, 034508 (2010), [arXiv:1004.4930].

\bibitem{Bulava:2010yg}
J.~Bulava {\em et~al.},
\newblock Phys. Rev. D {\bf 82}, 014507 (2010), [arXiv:1004.5072].

\bibitem{Engel:2010my}
G.~P. Engel, C.~B. Lang, M.~Limmer, D.~Mohler and A.~Sch{\"a}fer,
\newblock Phys. Rev. D {\bf 82}, 034505 (2010), [arXiv:1005.1748].

\bibitem{Engel:2012qp}
G.~P. Engel, C.~B. Lang and A.~Sch{\"a}fer,
\newblock Low-lying lambda baryons from the lattice,
\newblock 2012, [arXiv:1212.2032].

\bibitem{Mahbub:2012ri}
M.~S. Mahbub, W. Kamleh, Waseem, D.~B. Leinweber, P.~J. Moran
and A.~G. Williams,
\newblock Low-lying odd-parity states of the nucleon in lattice QCD,
\newblock 2012, [arXiv:1209.0240].

\bibitem{Lang:2011mn}
C.~B. Lang, D.~Mohler, S.~Prelovsek and M.~Vidmar,
\newblock Phys. Rev. D {\bf 84}, 054503 (2011), [arXiv:1105.5636].

\bibitem{Aoki:2011yj}
S.~Aoki {\em et~al.},
\newblock Phys. Rev. D {\bf 84}, 094505 (2011), [arXiv:1106.5365].

\bibitem{Feng:2010es}
X.~Feng, K.~Jansen and D.~B. Renner,
\newblock Phys. Rev. D {\bf 83}, 094505 (2011), [arXiv:1011.5288].

\bibitem{Lang:2012sv}
C.~B. Lang, L.~Leskovec, D.~Mohler and S.~Prelovsek,
\newblock Phys. Rev. D {\bf 86}, 054508 (2012), [arXiv:1207.3204].

\bibitem{Mohler:2012na}
D.~Mohler, S.~Prelovsek and R.~M. Woloshyn,
\newblock Phys. Rev. D {\bf 87}, 034501 (2013), [arXiv:1208.4059].

\bibitem{Pelissier:2012pi}
C.~Pelissier and A.~Alexandru,
\newblock {Resonance parameters of the $\rho$-meson from asymmetrical lattices},
\newblock 2012, [arXiv:1211.0092].

\bibitem{Dudek:2012xn}
J.~J. Dudek, R.~G. Edwards and C.~E. Thomas,
\newblock {Energy dependence of the $\rho$ resonance in ${\pi}{\pi}$ elastic
scattering from lattice QCD},
\newblock 2012, [arXiv:1212.0830]

\bibitem{Luscher:1985dn}
M.~L{\"u}scher,
\newblock Commun. Math. Phys. {\bf 104}, 177 (1986).

\bibitem{Luscher:1986pf}
M.~L{\"u}scher,
\newblock Commun. Math. Phys. {\bf 105}, 153 (1986).

\bibitem{Luscher:1990ux}
M.~L{\"u}scher,
\newblock Nucl. Phys. B {\bf 354}, 531 (1991).

\bibitem{Luscher:1991cf}
M.~L{\"u}scher,
\newblock Nucl. Phys. B {\bf 364}, 237 (1991).

\bibitem{Rummukainen:1995vs}
K.~Rummukainen and S.~Gottlieb,
\newblock Nucl. Phys. B {\bf 450}, 397 (1995), [arXiv:hep-lat/9503028].

\bibitem{Kim:2005gf}
C.~H. Kim, C.~T. Sachrajda and S.~R. Sharpe,
\newblock Nucl. Phys. {\bf B727}, 218 (2005), [arXiv:hep-lat/0507006].

\bibitem{Fu:2011xz}
Z.~Fu,
\newblock Phys. Rev. D {\bf 85}, 014506 (2012), [arXiv:1110.0319].

\bibitem{Leskovec:2012gb}
L.~Leskovec and S.~Prelovsek,
\newblock Phys. Rev. D {\bf 85}, 114507 (2012), [arXiv:1202.2145].

\bibitem{Gockeler:2012yj}
M.~G{\"o}ckeler {\em et~al.},
\newblock Phys. Rev. D {\bf 86}, 094513 (2012), [arXiv:1206.4141].

\bibitem{Bernard:2008ax}
V.~Bernard, M.~Lage, U.-G. Mei{\ss}ner and A.~Rusetsky,
\newblock JHEP {\bf 08}, 024 (2008), [arXiv:0806.4495].

\bibitem{Doring:2011ip}
M.~D{\"o}ring, J.~Haidenbauer, U.-G. Mei{\ss}ner and A.~Rusetsky,
\newblock Eur. Phys. J. {\bf A47}, 163 (2011), [arXiv:1108.0676].

\bibitem{Roca:2012rx}
L.~Roca and E.~Oset,
\newblock Phys.Rev. {\bf D85}, 054507 (2012), [arXiv:1201.0438].

\bibitem{Hall:2012wz}
J.~M.~M. Hall, A.~C.~P. Hsu, D.~B. Leinweber, A.~W. Thomas and R.~D. Young,
\newblock PoS {\bf LAT2012}, 145 (2012), [arXiv:1207.3562].

\bibitem{Meissner:2010rq}
U.~G. Mei{\ss}ner, K.~Polejaeva and A.~Rusetsky,
\newblock Nucl. Phys. B {\bf 846}, 1 (2011), [arXiv:1007.0860].

\bibitem{Giudice:2012tg}
P.~Giudice, D.~McManus and M.~Peardon,
\newblock Phys.Rev. {\bf D86}, 074516 (2012), [arXiv:1204.2745].

\bibitem{Peardon:2009gh}
Hadron Spectrum Collaboration, M.~Peardon {\em et~al.},
\newblock Phys. Rev. D {\bf 80}, 054506 (2009), [arXiv:0905.2160].

\bibitem{Michael:1985ne}
C.~Michael,
\newblock Nucl. Phys. B {\bf 259}, 58 (1985).

\bibitem{Luscher:1990ck}
M.~L{\"u}scher and U.~Wolff,
\newblock Nucl. Phys. B {\bf 339}, 222 (1990).

\bibitem{Blossier:2009kd}
B.~Blossier, M.~DellaMorte, G.~von Hippel, T.~Mendes and R.~Sommer,
\newblock JHEP {\bf 0904}, 094 (2009), [arXiv:0902.1265].

\bibitem{Hasenfratz:2008ce}
A.~Hasenfratz, R.~Hoffmann and S.~Schaefer,
\newblock Phys. Rev. D {\bf 78}, 054511 (2008), [arXiv:0806.4586].

\bibitem{Hasenfratz:2008fg}
A.~Hasenfratz, R.~Hoffmann and S.~Schaefer,
\newblock Phys. Rev. D {\bf 78}, 014515 (2008), [arXiv:0805.2369].

\bibitem{Arndt:2006bf}
R.~A. Arndt, W.~J. Briscoe, I.~I. Strakovsky and R.~L. Workman,
\newblock Phys. Rev. C {\bf 74}, 045205 (2006), [arXiv:nucl-th/0605082].

\bibitem{Manley:1992yb}
D.~Manley and E.~Saleski,
\newblock Phys. Rev. D {\bf 45}, 4002 (1992).

\bibitem{Koch:1985bn}
R.~Koch,
\newblock Nucl. Phys. A {\bf 448}, 707 (1986).

\bibitem{Cutkosky:1979fy}
R.~E. Cutkosky, C.~P. Forsyth, R.~E. Hendrick and R.~L. Kelly,
\newblock Phys. Rev. D {\bf 20}, 2839 (1979).

\bibitem{Beringer:1900zz}
Particle Data Group, J.~Beringer {\em et~al.},
\newblock Phys. Rev. D {\bf 86}, 010001 (2012).

\bibitem{McNeile:2007fu}
C.~McNeile,
\newblock PoS {\bf LATTICE2007}, 019 (2007), [arXiv:0710.0985].

\bibitem{Jansen:2008wv}
K.~Jansen, C.~Michael and C.~Urbach,
\newblock Eur. Phys. J. C {\bf 58}, 261 (2008), [arXiv:0804.3871].

\bibitem{Dudek:2011tt}
J.~J. Dudek {\em et~al.},
\newblock Phys. Rev. D {\bf 83}, 111502 (2011), [arXiv:1102.4299].

\bibitem{Liu:2012ze}
Hadron Spectrum Collaboration, L.~Liu {\em et~al.},
\newblock JHEP {\bf 1207}, 126 (2012), [arXiv:1204.5425].

\bibitem{Mahbub:2010rm}
M.~S. Mahbub, W.~Kamleh, D.~B. Leinweber, P.~J. Moran and A.~G. Williams,
\newblock Phys.Lett. {\bf B707}, 389 (2012), [arXiv:1011.5724].

\bibitem{Engel:2013ig}
G.~P. Engel, C. B. Lang, D. Mohler and A. Sch{\"a}fer,
\newblock QCD with two light dynamical Chirally Improved quarks: baryons,
\newblock 2013, [arXiv:1301.4318].

\bibitem{Weinberg:1966kf}
S.~Weinberg,
\newblock Phys. Rev. Lett. {\bf 17}, 616 (1966).

\bibitem{Tomozawa:1966jm}
Y.~Tomozawa,
\newblock Nuovo Cim. {\bf A46}, 707 (1966).

\end{thebibliography}
\end{document}